\documentclass[a4paper,11pt]{article}

\usepackage{jinstpub} 

\usepackage{lscape} 

\usepackage{graphicx}
\usepackage{dcolumn}
\usepackage{amsthm}
\usepackage{bm}
\usepackage{color}
\usepackage{amsmath}
\usepackage{amsbsy}
 \usepackage{microtype}

\usepackage{bbm}
\usepackage{hyperref}
\usepackage{todonotes}
\usepackage{hhline}
\hypersetup{
    colorlinks = true
}
\usepackage{textcomp}
\usepackage{etoolbox}
\usepackage{enumitem}
\usepackage{mathrsfs}
\usepackage{bbm}
\usepackage[ruled,vlined]{algorithm2e}
\usepackage{algorithmic}
\usepackage{amsmath}

\usepackage{hyperref}
\usepackage{subfig}
\usepackage{booktabs, tabularx}
\usepackage{array}
\usepackage{cellspace}
\usepackage{natbib}
\newcolumntype{L}[1]{>{\raggedright\arraybackslash}p{#1}}
\newcolumntype{Y}{>{\raggedright\arraybackslash}X} 
\setenumerate[0]{label=$(\alph*)$}
\renewrobustcmd{\bfseries}{\fontseries{b}\selectfont}
\renewrobustcmd{\boldmath}{}
\newrobustcmd{\B}{\bfseries}

\theoremstyle{definition}

\theoremstyle{remark}

\numberwithin{equation}{section}

\newcommand{\norm}[1]{\left\lVert#1\right\rVert}

\newcommand{\bbR}{\mathbb{R}}

\title{\boldmath Deep Neural Network Uncertainty Quantification for LArTPC Reconstruction}

\author[a,1]{D. Koh,\note{Corresponding author.}}
\author[c]{A. Mishra}
\author[c]{and K. Terao}

\affiliation[a]{Stanford University,\\450 Serra Mall, Stanford, United States}
\affiliation[c]{SLAC National Accelerator Laboratory,\\2575 Sand Hill Rd, Menlo Park, United States}

\emailAdd{koh0207@stanford.edu}

\abstract{We evaluate uncertainty quantification (UQ) methods for deep learning applied to liquid argon time projection chamber (LArTPC) physics analysis tasks. As deep learning applications enter widespread usage among physics data analysis, neural networks with reliable estimates of prediction uncertainty and robust performance against overconfidence and out-of-distribution (OOD) samples are critical for their full deployment in analyzing experimental data. While numerous UQ methods have been tested on simple datasets, performance evaluations for more complex tasks and datasets are scarce. We assess the application of selected deep learning UQ methods on the task of particle classification using the PiLArNet \cite{pilarnet} monte carlo 3D LArTPC point cloud dataset. We observe that UQ methods not only allow for better rejection of prediction mistakes and OOD detection, but also generally achieve higher overall accuracy across different task settings. We assess the precision of uncertainty quantification using different evaluation metrics, such as distributional separation of prediction entropy across correctly and incorrectly identified samples, receiver operating characteristic curves (ROCs), and expected calibration error from observed empirical accuracy. We conclude that ensembling methods can obtain well calibrated classification probabilities and generally perform better than other existing methods in deep learning UQ literature. }

\begin{document}
\maketitle
\flushbottom


\section{Introduction}


Deep learning has largely established itself as a dominant method in machine learning applications, in part due to its competence at a variety of well-known tasks such as image recognition, natural language processing, and automated control applications. As such, scientists in both artificial intelligence and the physical sciences have been investigating ways to realize deep learning's success in more complex domains of fundamental research. The trend for integrating deep learning for physics data reconstruction has been particularly notable in experimental particle physics, where large data generation from particle detectors such as liquid argon time projection chambers (LArTPCs) and the Large Hadron Collider (LHC) naturally prepare fertile grounds for deep learning models. 

Using deep learning for fundamental research, however, presents complications that are often omitted in many common industrial use cases, where practitioners generally attend to achieving state-of-the-art results with respect to a family of conventional 
performance metrics. In particular, one of the most pressing issues with using deep neural networks for fundamental research is developing robust and consistent methods for quantifying uncertainties of their predictions. Deep neural networks are unable to recognize out-of-distribution examples and habitually make incorrect predictions with high confidence for such cases \cite{nguyen2015deep, hendrycks2016baseline}. Uncertainty in predictions has had serious consequences while applying deep learning to high-regret and safety-critical applications such as automated driving \cite{NTSB0, NTSB1, NTSB2}, law enforcement \cite{dodds2018}, medical sciences\cite{sundararajan2017axiomatic}, etc. Overconfidence for out-of-distribution examples also emphasizes the need for deep learning models to indicate whether a given prediction is to be trusted or not. Undoubtedly, for deep neural nets to be integrated into the physics measurement process, such characteristics of deterministic neural networks must be addressed by an effective method for uncertainty quantification (UQ). 

As the need for UQ gradually escalated in domains such as autonomous driving and medicine, UQ methods diversified into a variety of different approaches under the name of Bayesian Deep Learning (BDL) \cite{wilson2020bayesian}, but with scarce substantial application in the physical sciences. Moreover, most BDL methods have been benchmarked on simplified datasets (MNIST\cite{deng2012mnist}, CIFAR10\cite{Krizhevsky09learningmultiple}), which are not representative of the complexity of physics data reconstruction process. Modern accelerator neutrino experiments such as ICARUS\cite{rubbia2011underground} and DUNE\cite{dune, dune2016long} offer ideal grounds for testing the efficacy of BDL in UQ, due to its recent adaptation and the moderate success of deep learning based reconstruction techniques. The benefit derived from a detailed assessment of different UQ algorithms on a complex, multi-objective task such as LArTPC data reconstruction is two-fold: allow practitioners in machine learning to evaluate BDL's applicability in a real-world setting \cite{nado2021uncertainty} and enable physicists to design neural network that produce well justified uncertainty estimates for rejecting erroneous predictions and detecting out-of-distribution instances\cite{psaros2023uncertainty,mishra2021uncertainty}. 

Practitioners of deep learning in LArTPC reconstruction agree on the need for calibrated uncertainty bounds for deep learning model predictions along with OOD robustness. However, numerous different uncertainty quantification algorithms have been proposed for deep learning. These range from empirical approaches (such as bootstrapped ensembles\cite{efron1994introduction, deepensembles}), Bayesian approaches (such as EDL\cite{edl-class}, HMC\cite{neal2012bayesian}) and hybrid approaches (such as MC Dropout\cite{mcdropout}). None of these have been tested for complex applications such as LArTPC reconstruction. In this investigation, we select the most promising uncertainty quantification approaches from each of these categories, test and evaluate them with respect to critical intermediate reconstruction tasks: particle classification and particle shape (semantic) segmentation. For a given 3D LArTPC image, particle classification is defined to be the task of predicting the type--the Particle Data Group (PDG) code--of the particle that has created the energy depositions recorded by the detector. Particle shape segmentation, which we will refer to as semantic segmentation\cite{laura, guo2018review}, is the process of predicting a category for every non-zero pixel of the LArTPC image. For instance, in a study by the MicroBooNE collaboration \cite{microboone_segmentation}, a neural network was trained to differentiate electromagnetic showers and tracks for each pixel. We briefly summarize the different methodologies and discuss the apparent advantages and disadvantages of using each of the proposed models in the following section. We describe in detail the monte-carlo generated 3D LArTPC particle image dataset and state any assumptions or additional information that was used to train and evaluate each model. In Section IV, we present quantitative performance evaluation of different UQ models on three different settings of single particle classification, multi-particle classification, and semantic segmentation, using a variety of quantitative metrics to measure UQ fidelity. 

While quantifying the predictive uncertainty for machine learning models, 
there are two broad classes of uncertainty that need to be considered: 
\textit{epistemic} uncertainty and \textit{aleatoric} uncertainty. \textit{Epistemic} uncertainty \cite{smith2013uncertainty} arises due to a lack of knowledge regarding the dynamics of the system under consideration, or an inability to express the underlying dynamics using models. It is also referred to as \textit{reducible} uncertainty, as one could in principle reduce it by collecting more data to further constrain the model. Quantifying epistemic uncertainties is essential for safety critical applications, for instance to understand and identify examples that are significantly different from the training data \cite{kendall2017uncertainties}. Epistemic uncertainties are 
important while learning from small datasets where the training data is 
sparse, globally or locally. Epistemic uncertainty can be further divided into 
\textit{parameter} and \textit{structural} uncertainty. Parameter uncertainty\cite{blundell2015weight}
captures our ignorance about the exact combination of model coefficients 
that generated the training data, for example the weights and biases of a neural network model, and may be minimized with larger corpora of training data. Structural uncertainty in deep learning \cite{antoran2020depth} captures our inability to represent the underlying process via a limited model. In the context of deep learning, it represents the uncertainties in the model architecture and hyperparameters like the activation functions utilized. 

\begin{figure}[!bth!]
\begin{center}
\includegraphics[width=1.05\textwidth]{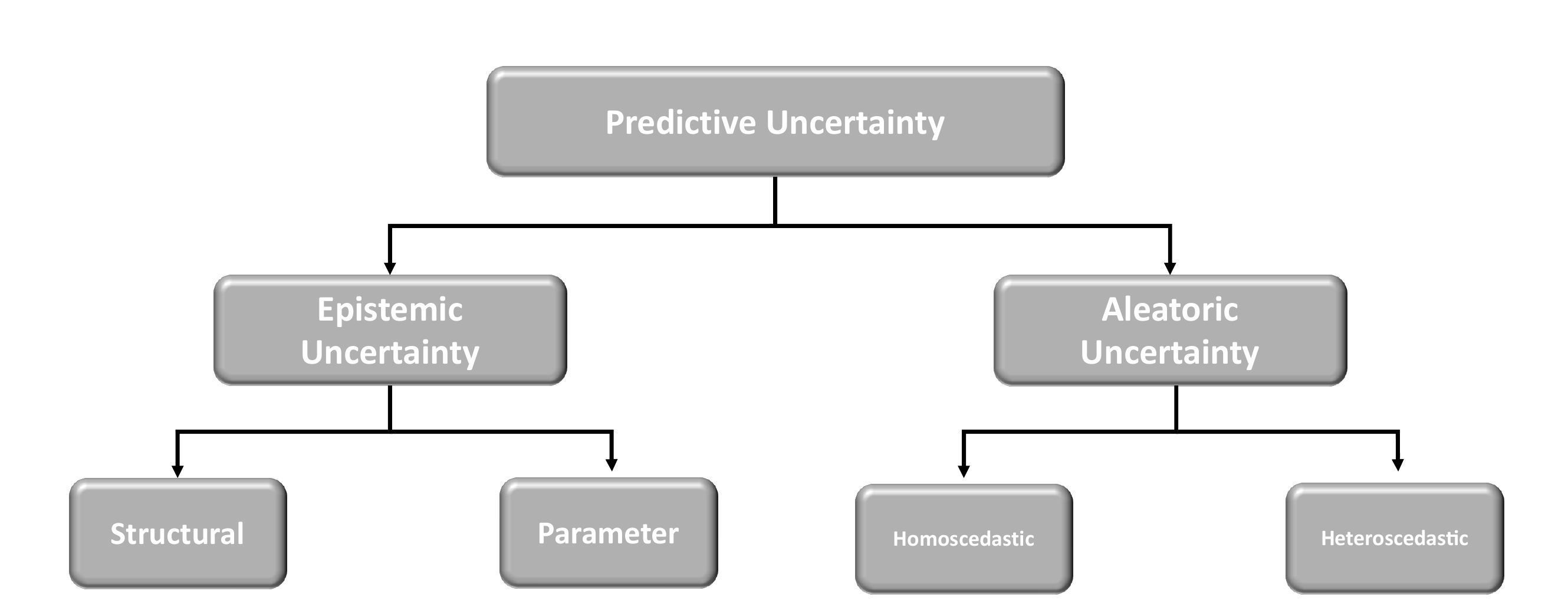}
 \caption{\label{fig:UQ_Flowchart}Schematic outlining the types of uncertainty in Deep Learning model predictions.}
\end{center}
\end{figure}

Aleatoric uncertainty \cite{smith2013uncertainty} is often referred to as 
irreducible uncertainty or stochastic uncertainty. In scientific 
applications, this arises due to noise in the training data, projection of data 
onto a lower dimensional subspace, absence of important input features in the 
data, etc. While large datasets can ameliorate epistemic uncertainty, 
aleatoric uncertainties are immune to such remedies and thus often named as \textit{irreducible} uncertainties. Aleatoric uncertainties\cite{kendall2017uncertainties} 
can be divided into \textit{homoscedastic} and \textit{heteroscedastic}
uncertainty. Homoscedastic uncertainties are independent of the input data, 
but depend on the application. In contrast, heteroscedastic uncertainties 
strongly depend on the model input. These exhibit variability across 
examples with some inputs having significantly higher noise in the 
predictions for some samples than others. 

These two classes of uncertainty, epistemic and aleatoric, are significantly 
different in their nature and in the manner that they are dealt with in 
uncertainty quantification. For instance, Aleatoric uncertainties arise due to 
random variability and thus, introduce random variance in the model 
predictions. Epistemic uncertainties on the other hand arise due to 
incomplete representation of underlying physics and add a bias to the 
simulation results. This broad division regarding the sources and nature of predictive uncertainties is schematically outlined in Figure \ref{fig:UQ_Flowchart}. 

In physical sciences, the divisions between different kinds of uncertainty and their definitions are different. By precedent, uncertainty in physical sciences is often demarcated into statistical and systematic uncertainties. Statistical uncertainties refer to errors that may be quantified by statistical analysis over a series of measurements. Systematic uncertainties occur due to limitations of models and theory, and thus cannot be treated the same as statistical. As a mapping between the nomenclature utilized in the different fields, epistemic uncertainties are always systematic uncertainties, and statistical uncertainties are always aleatoric. 

\section{Methods of Uncertainty Quantification in Deep Learning}


In machine learning, a given dataset is divided into \textit{training} and \textit{test} sets. The \textit{training} set is used to find the optimal model that best describes the data, with respect to some numerical criterion. The \textit{test} set is later used to estimate the generalization error and select the best performing trained model, which has an inherent bias towards the \textit{training} set. The generalization error is the expected error of the model on data that it has not been trained on. 

We first introduce mathematical notations used throughout the paper. Let $X = \{x_1, x_2, ..., x_N \}$ and $Y = \{y_1, y_2, ..., y_N\}$ be the input data and labels in the training set, and let $\hat{X} = \{ {\hat{x}}_1, {\hat{x}}_2, ..., {\hat{x}}_M \}$ and $\hat{Y} = \{ {\hat{y}}_1, {\hat{y}}_2, ..., {\hat{y}}_M \}$ denote the corresponding quantities in the test set. Uncertainty quantification for $C$-class classification may be defined as the process of determining a posterior distribution $\mathbf{p}(\hat{x} \mid X, Y) \in S_C$, where $S_C$ is the $C$-dimensional unit simplex. Among numerous models and studies focusing on the uncertainty quantification of neural networks \cite{uqsurvey1, bdlsurvey}, we focus on methods designed for multi-class classification tasks that require minimal changes to popular neural network architectures. In this paper, we consider three specific UQ methods: model ensembling \cite{deepensembles}, Monte Carlo Dropout (MCD) \cite{gal2016}, and Evidential Deep Learning (EDL) \cite{edl-class, edl-reg}. 

\subsection{Deterministic Neural Networks}

Let $f_\theta$ refer to a neural network with trainable parameters $\theta$. Since each prediction of the network depends on the the training set data and label pair $(X, Y)$, we denote the model predictions as $\mathbf{f}_\theta(\hat{x}; X,Y) \in \bbR^C$ with $j \in \{1, 2, ..., M\}$, where $\hat{x}$ is an example in the test set and $C$ is the number of distinct labels in our dataset. 
\begin{equation}
\mathbf{f}_\theta(\hat{x}; X, Y) \equiv \begin{bmatrix}
    f_\theta(\hat{x}; X,Y)_1 \\
    f_\theta(\hat{x}; X,Y)_2 \\
    \vdots \\
    f_\theta(\hat{x}; X,Y)_C
\end{bmatrix} \equiv \begin{bmatrix}
    \hat{z}_1 \\
    \hat{z}_2 \\
    \vdots \\
    \hat{z}_C
\end{bmatrix} \in \bbR^C. 
\end{equation}
Here, it is typical to call each predicted score value $\hat{z}_c \equiv f_\theta(\hat{x}; X, Y)_c \in \bbR$ as \textit{logits}. The logit value of each class may be interpreted as the log-odds value of predicting the given example $\hat{x}$ as class $c \in C$. In mathematical terms, the logit function, $L(p)$ maps probabilities to their log-odds, $L(p)=ln(\frac{p}{1-p})$. The predicted label $\hat{y}$ corresponding to $\hat{x}$ is then given as:
\begin{equation}
    \hat{y} \equiv \underset{{c \in C}}{\text{argmax}} [\hat{z}_c]. 
\end{equation}

We may also apply the softmax function to the logit value to obtain the predicted probabilities $\hat{p}_c$:
\begin{equation}
    \mathbf{p}(\hat{x} \mid X,Y) \equiv \begin{bmatrix} \frac{e^{\hat{z}_1}}{ \underset{c \in C}{\sum} e^{\hat{z}_c} } &

    \frac{e^{\hat{z}_2}}{ \underset{c \in C}{\sum} e^{\hat{z}_c} } &

    \cdots &

    \frac{e^{\hat{z}_C}}{ \underset{c \in C}{\sum} e^{\hat{z}_c} }

    \end{bmatrix} \equiv \begin{bmatrix}
    \hat{p}_1 & \hat{p}_2 & \cdots & \hat{p}_C
    \end{bmatrix}. 
\end{equation}

For deterministic networks, the posterior is simply given by the predicted probabilities $\hat{p}_1, ..., \hat{p}_C$. It is well known that for deterministic neural networks, the probability values $\hat{p}_c \in [0,1]$ cannot be interpreted as valid estimate of the empirical accuracy of predicting a given example as class $c$ and are often denoted as being \textit{uncalibrated}. For example, in a binary classification task with input $\hat{x}$, $\mathbf{p}(\hat{x} \mid X,Y) \in [0,1]$, and $\hat{y} \in \{0, 1\}$, we may ask if the neural network's prediction $\mathbf{\hat{p}} \equiv \mathbf{p}(\hat{x} \mid X,Y) = 0.7$ corresponds to a 70\% chance of classifying $\hat{x}$ as $\hat{y} = 1$. UQ methods for neural networks may be summarized as attempts to estimate the true empirical accuracy from neural network outputs such as (but not limited to) $\mathbf{p}(\hat{x} \mid X,Y)$. 


\subsection{Ensembling Methods}

In the context of deep learning, model ensembling refers to the method of training multiple instances of the same architecture with different random initialization seeds. In Naive Ensembling (NE), one trains each member of the ensemble on the same training dataset, resulting in $N$ networks with identical architecture but different parameter values. The posterior distribution for ensembling methods is obtained by computing the average of deterministic posteriors $\mathbf{p}_t(\hat{x} \mid X,Y)$:

\begin{equation}
    \mathbf{p}(\hat{x}\mid X, Y) \approx \frac{1}{N} \sum_{i = 1}^N \mathbf{p}_i(\hat{x}\mid X, Y), \quad \mathbf{p}_i(\hat{x}\mid X,Y) \equiv \begin{bmatrix}
        f^{(i)}_{\theta}(\hat{x}; X,Y)_1 \\
        f^{(i)}_{\theta}(\hat{x}; X,Y)_2 \\ 
        \vdots \\
        f^{(i)}_{\theta}(\hat{x}; X,Y)_C
    \end{bmatrix}.
\end{equation}

Often, to achieve better generalization and stability, Bootstrapped \cite{mooney1993bootstrapping} Ensembling (BE) (or bagging) is preferred over naive ensembling. This is done by training each ensemble member on a dataset reorganized by sampling $n$ examples from the full training set with replacement. If the size of the resampled dataset is equal to that of the original training set, each ensemble member is expected to see approximately 63\% of the original training set. For classification, it is standard to use the most common label among the ensemble members as the final prediction, while for regression one usually computes the empirical mean. When an ensemble consists of a collection of neural networks trained with respect to a \textit{proper scoring rule} \cite{properscoringrules} and often coupled with an optional adversarial training routine, the ensemble is termed as a\textit{deep ensemble} \cite{deepensembles}. 

Ensemble methods are the one of the simplest UQ methods that require no additional changes to the underlying model architecture, although the high computational cost in training $M$ architecturally identical models and performing $M$ forward passes for each prediction often renders them impractical for some memory or time consuming tasks. 

\subsection{Monte Carlo Dropout}

Monte Carlo Dropout is a Bayesian technique introduced in \cite{gal2016}, where one approximates the network's posterior distribution of class predictions by collecting samples obtained from multiple forward passes of dropout regularized networks. \textit{Dropout regularization} \cite{dropout} involves random omissions of hidden features during train time, which is equivalent to masking rows of weight matrices. Inclusion of dropout layers mitigates model overfitting and is empirically known to improve model accuracy \cite{dropout}. A key observation of \cite{gal2016} is that under suitable assumptions on the bayesian neural network prior and training procedure, sampling $T$ predictions from the BNN's posterior is equivalent to performing $T$ stochastic forward passes with dropout layers fully activated. In this manner, the full posterior distribution may be approximated by monte-carlo integration of the posterior softmax probability vector $\mathbf{p}(x^*; X, Y)$:
\begin{equation}
    \mathbf{p}(\hat{x}\mid X, Y) \approx \frac{1}{T} \sum_{t = 1}^T \mathbf{p}_t(\hat{x}\mid X, Y), \quad \mathbf{p}_t(\hat{x}\mid X,Y) \equiv \begin{bmatrix}
        f^{(t)}_{\theta}(\hat{x}\mid X,Y)^1 \\
        f^{(t)}_{\theta}(\hat{x}\mid X,Y)^2 \\ 
        \vdots \\
        f^{(t)}_{\theta}(\hat{x}\mid X,Y)^C
    \end{bmatrix}.
\end{equation}
where $T$ denotes the number of stochastic forward passes. In MCDropout, each output $f_\theta^{(t)}$ from different forward passes $t$ share the same model weights $\theta$ but have different hidden features of the network omitted. This way, each prediction from separate stochastic forward passes effectively uses a slightly different set of hidden features for computing the final probability values $\mathbf{p}_t$. As with ensembling methods, the final prediction of MCDropout for classification is given by the majority vote among all stochastic forward passes. For regression, we again compute the empirical mean. As evident from the apparent similarities, MCDropout networks may also be interpreted as a form of ensemble learning \cite{dropout}, where each stochastic forward pass corresponds to a different realization of a trained neural network. 

Implementing MCDropout requires one to modify the underlying neural network architecture to include dropout layers and configuring them to behave stochastically during test time. Often the location of dropout layers can critically affect prediction performance, and for convolutional neural networks the decision is made via trial-and-error \cite{kendall2017uncertainties}. Also, for memory intensive tasks such as semantic segmentation, sample collection by multiple forward passes can accumulate rapidly towards high computational cost, similar to ensembling methods. 

\subsection{Evidential Deep Learning}

Evidential Deep Learning (EDL) \cite{edl-class, edl-reg}, refers to a class of deep neural networks that exploit conjugate prior relationships to model the posterior distribution analytically. For multi-class classification, the distribution over the space of all probability vectors $\mathbf{p} = (p_1, ..., p_c)$ is modeled by a Dirichlet distribution with $c$ concentration parameters $\boldsymbol{\alpha} = \alpha_1, ..., \alpha_c$:
\begin{equation}
    D(\mathbf{p} \mid \boldsymbol{\alpha}) = \frac{1}{B(\boldsymbol{\alpha})} \prod_{i = 1}^c p_i^{\alpha_i - 1},
\end{equation}
where $\alpha_i \geq 1$ for all $i$, $B( \cdot )$ denotes the $c$-dimensional multivariate Beta function, and $\mathbf{p}$ is in the $c$-unit simplex $\mathcal{S}_c$:
\begin{equation}
    \mathcal{S}_c = \{\mathbf{v} \in \mathbb{R}^c : \sum_{i=1}^c v_i = 1\}.
\end{equation}

In contrast to deterministic classification neural networks that minimize the cross-entropy loss by predicting the class logits, evidential classification networks predict the concentration parameters $\boldsymbol{\alpha}$. The expected value of the $k$-th class probability under the distribution $D(\mathbf{p} \mid \boldsymbol{\alpha})$ is then given analytically as 
\begin{equation}
    \hat{p}_k = \frac{\alpha_k}{\sum_{i = 1}^c \alpha_i}.
\end{equation}

To estimate the concentration parameters, several distinct loss functions are available. Since in practice the concentration parameters are obtained from neural network predictions, we denote $\boldsymbol{\alpha}_\theta$ as the predicted concentration parameters of an EDL model. The \textit{marginal likelihood loss} (MLL) is given by:
\begin{equation}
    \mathcal{L}_{MLL}(\theta) = -\log{\left( \int \prod_{i=1}^c p_i^{y_i} D(\mathbf{p} \mid \boldsymbol{\alpha}_\theta) \ d\mathbf{p} \right)}.
\end{equation}

The \textit{Bayes risk} (posterior expectation of the risk) of the \textit{log-likelihood} (BR-L) formulation yields:
\begin{equation}
    \mathcal{L}_{BR}(\theta) = \int \left[ \sum_{i=1}^c -y \log{(p_i)} \right] D(\mathbf{p} \mid \boldsymbol{\alpha}_\theta) \ d\mathbf{p}. 
\end{equation}

The \textit{Bayes risk} of the \textit{Brier score} (BR-B) may also be used as an alternative optimization objective:
\begin{equation}
    \mathcal{L}_{BS}(\theta) = \int \norm{\mathbf{y} - \mathbf{p}}_2^2 \ D(\mathbf{p} \mid \boldsymbol{\alpha}_\theta) \ d\mathbf{p}. 
\end{equation}

From Sensoy et. al. \cite{edl-class}, analytic integration of the aforementioned loss functions gives closed form expressions that are suited for gradient based optimization of the parameters $\theta$. 

EDL methods have the immediate advantage of requiring only one single pass to access the full posterior distribution, at a price of restricting the space of posterior functions onto the appropriate conjugate prior forms. Also, EDL methods only require one to modify the loss function and the final layer of its deterministic baseline (if necessary), which allows flexible integration with complex, hierarchical deep neural architectures similar to the full LArTPC reconstruction chain. 
However, as we later observe, EDL methods generally fall short on various UQ evaluation metrics compared to ensembling and MCDropout, depending on task specifics. 

\section{Evaluating Uncertainty Quantification Methods}


As stated in \cite{deepensembles}, the goal for uncertainty quantification in deep learning is two-fold: to achieve better alignment of estimated probability with their long-run empirical accuracy and to serve as mis-classification or out-of-distribution alarms that could be used for rejecting erroneous predictions. 
The first condition, which we term \textit{calibration fidelity}, may be evaluated by plotting the \textit{reliability diagrams} \cite{reliability_diagrams}, which are constructed by binning the predicted probabilities (often termed \textit{confidence}) into equal sized bins and plotting the bin centers in the $x$-axis and the empirical accuracy of the bin members in the $y$-axis. 

As calibration fidelity measurements using reliability diagrams are originally designed for binary classifiers, there have been numerous proposals for their extensions to multi-class classifiers \cite{guo_calibration, Widmann2019CalibrationTI, marginal_calibration}. We consider two relatively simple methods; the first is a standard used in Guo et. al. \cite{guo_calibration}, where only the predicted probability for the most confident prediction of each sample is used to plot the reliability diagram. We refer to this mode of assessment as \textit{max-confidence} calibration fidelity. An alternative method is to evaluate calibration for each of the $K$ classes separately, as in B. Zadrozny and C. Elkan \cite{marginal_calibration}. We refer to this mode as \textit{marginal} calibration fidelity. 

The closer the reliability diagram curve is to the diagonal, the better calibrated a given classifier is, in the sense of calibration fidelity. The deviation of a given classifier from the diagonal could be summarized by computing the \textit{adaptive calibration error} (ACE) \cite{ace} for each reliability diagram:
\begin{equation}
    ACE = \frac{1}{C} \frac{1}{R} \sum_{c=1}^C \sum_{r = 1}^R |\text{acc}(r, c) - \text{conf}(r, c)|. 
\end{equation},
where
\begin{equation}
    \text{acc}(r,c) = \frac{\text{\# of samples in bin $r$, classified as class $c$}}{\text{\# of samples in bin $r$}}
    \label{eq:acc}
\end{equation}
and
\begin{equation}
    \text{conf}(r,c) = \frac{\text{sum of predicted probabilities $p_j^c$ in bin $r$}} {\text{\# of samples in bin $r$}}.
    \label{eq:conf}
\end{equation}
In short, $\text{acc}(r,c)$ is the restricted empirical accuracy of samples with predicted probabilities between the boundaries of bin $r$, and $\text{conf}(r,c)$ is the average predicted probability for class $c$ within that same bin $r$. 
Here, $C$ is the number of unique classes and $R$ denotes the number of \textit{equal-sample} bins used to plot the reliability diagram of class $c$. 
In contrast to the \textit{expected calibration error} (ECE) \cite{ece} that uses \textit{equal-range} bins to compute the summary value, the ACE uses bins with equal number of samples. As such, ACE is better suited for comparing calibration fidelity for models with highly skewed probability distributions. Note that the ACE metric depends on how the reliability diagram is constructed: for definitions \eqref{eq:acc} and \eqref{eq:conf}, we use the \textit{marginal} reliability diagram. For \textit{max-confidence} diagrams, we define $\text{acc}(r)$ to be the empirical accuracy of the most confidence prediction and $\text{conf}(r)$ is the average maximum predicted probability of samples in bin $r$. 


Another metric of uncertainty quantification measures the model's \textit{discriminative capacity} to mis-classified or out-of-distribution samples. In practice, uncertainty quantification models have the capacity to reject predictions based on a numerical estimate of the trustworthiness of the prediction in question. For example, in a classification setting the entropy of the predicted softmax probability distribution (\textit{predictive entropy}) could be used as a measure of confusion, as entropy is maximized if the predictive distribution reduces to a uniform distribution over $K$ classes. In this construction, it is desirable to have the predicted entropy distributions of correctly and incorrectly classified samples to be as separated as possible. To compute the extent of distributional separation, we may use the first Wasserstein distance \cite{wasser} between the predictive entropy distributions:
\begin{equation}
    W_1(u,v) = \inf_{\pi \in \Gamma(u,v)} \int_{\bbR \times \bbR} |x-y| \ d\pi(x,y). 
\end{equation}
where $u$ and $v$ are two probability distributions, $\Gamma(u,v)$ is the set of all joint probability measures in $\bbR^2$. We use the Wasserstein distance with the $L_1$ metric due to its simple computational implementation \cite{wasser}. 

Sensitivity may also be measured by computing the area under the receiver operating characteristic curve (AUROC), also known as the concordance statistic (\textit{c}-statistic) \cite{tibshirani}. Using predictive entropy as the threshold value, the ROC curve is constructed by plotting the false positive rate (incorrect predictions) in the $x$-axis and the true positive rate (correct predictions) in the $y$-axis at different threshold levels. In this setting, the AUROC is the probability that a randomly chosen correct prediction will have a lower predictive entropy than that of a randomly chosen incorrect prediction \cite{roc_analysis}. 

\section{Datasets and Network Architectures}

\textbf{Dataset}: As previously stated, we use the PiLArNet \cite{pilarnet} public dataset for semantic segmentation and multi-particle classification. PiLArNet\cite{pilarnet} is a collection of simulated 2D/3D  images prepared for developing data analysis techniques for LArTPC experiments. In the first step, two custom algorithms--MultiParticleVertex (MPV) and MultiParticleRain (MPR)--generate a list of particles to be produced in each image. The MPV generator produces $N$ particles from a common 3D spatial point, while for the MPR generator, the particles may originate from different spatial points in the image. The MPV and MPR generator is designed to emulate neutrino interactions (MPV) and cosmic rays (MPR). Once each particle has been produced by MPV and MPR generators, we use the GEANT4 \cite{Geant4} particle tracking simulation to produce the energy depositions. 

For semantic segmentation and multi-particle classification, we use the 768px cubic PiLArNet dataset to train and test our models. For single-particle classification, we use a 1024px cubic dataset, which is generated using the same algorithm and parameters (apart from the difference in total resolution) as those of the PiLArNet 768px data. 

\begin{figure*}[htp]
    \centering
    \includegraphics[width=0.98\textwidth]{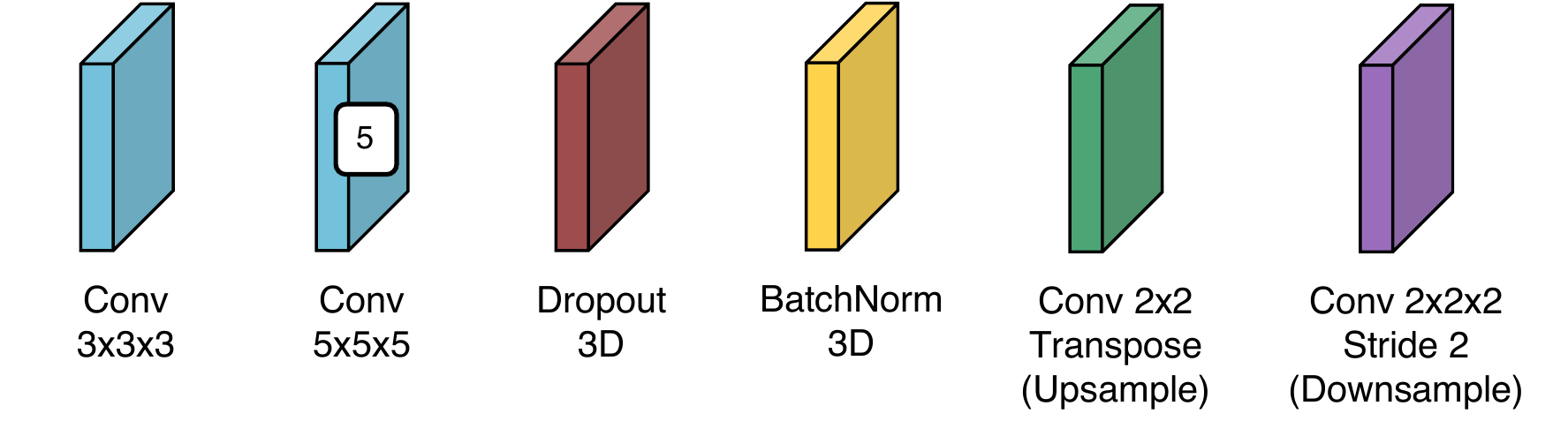}
    \caption{Sparse-CNN layer definitions for architecture reference.}
    \label{fig:arch1}
\end{figure*}

\begin{figure*}[htp]
    \centering
    \includegraphics[width=0.98\textwidth]{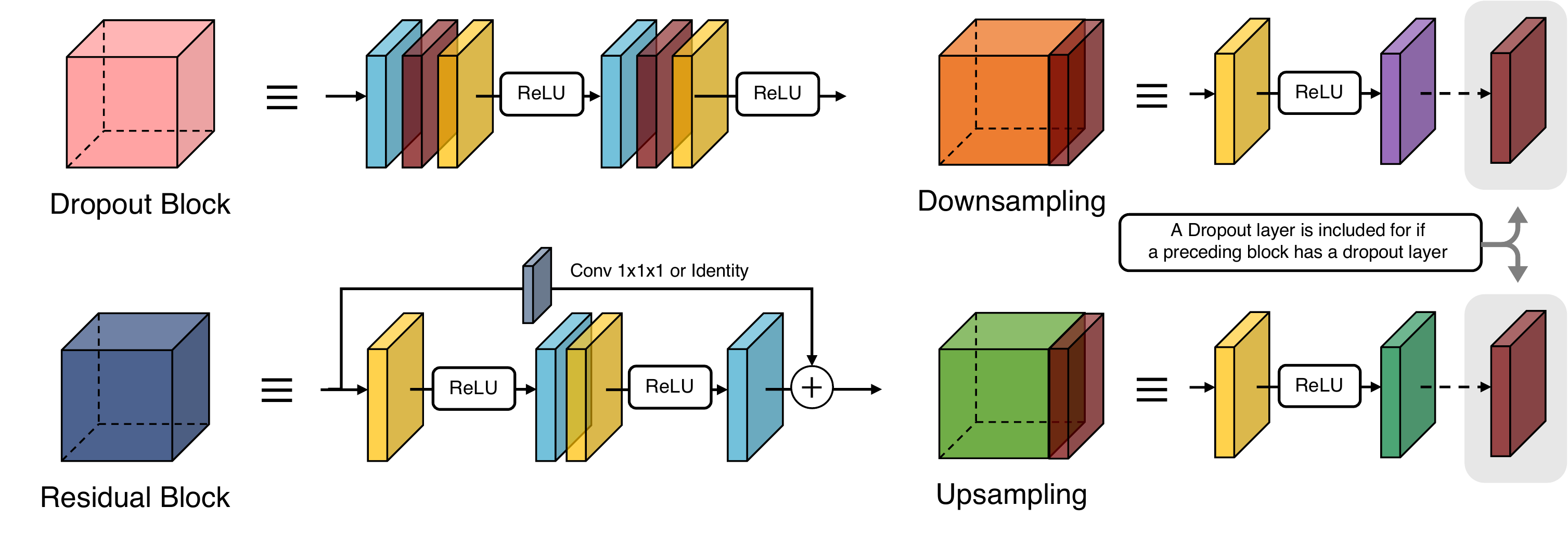}
    \caption{Sparse-CNN block definitions for architecture reference.}
    \label{fig:arch2}
\end{figure*}

\begin{figure*}[htp]
    \centering
    \includegraphics[width=0.98\textwidth]{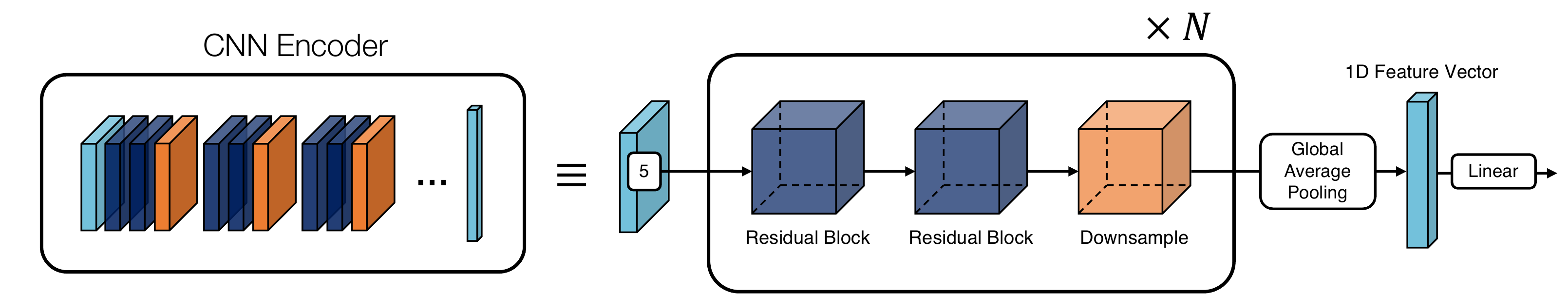}
    \caption{Sparse-CNN architecture for single particle classifiers.}
    \label{fig:arch3}
\end{figure*}

\begin{figure*}[htp]
    \centering
    \includegraphics[width=\textwidth]{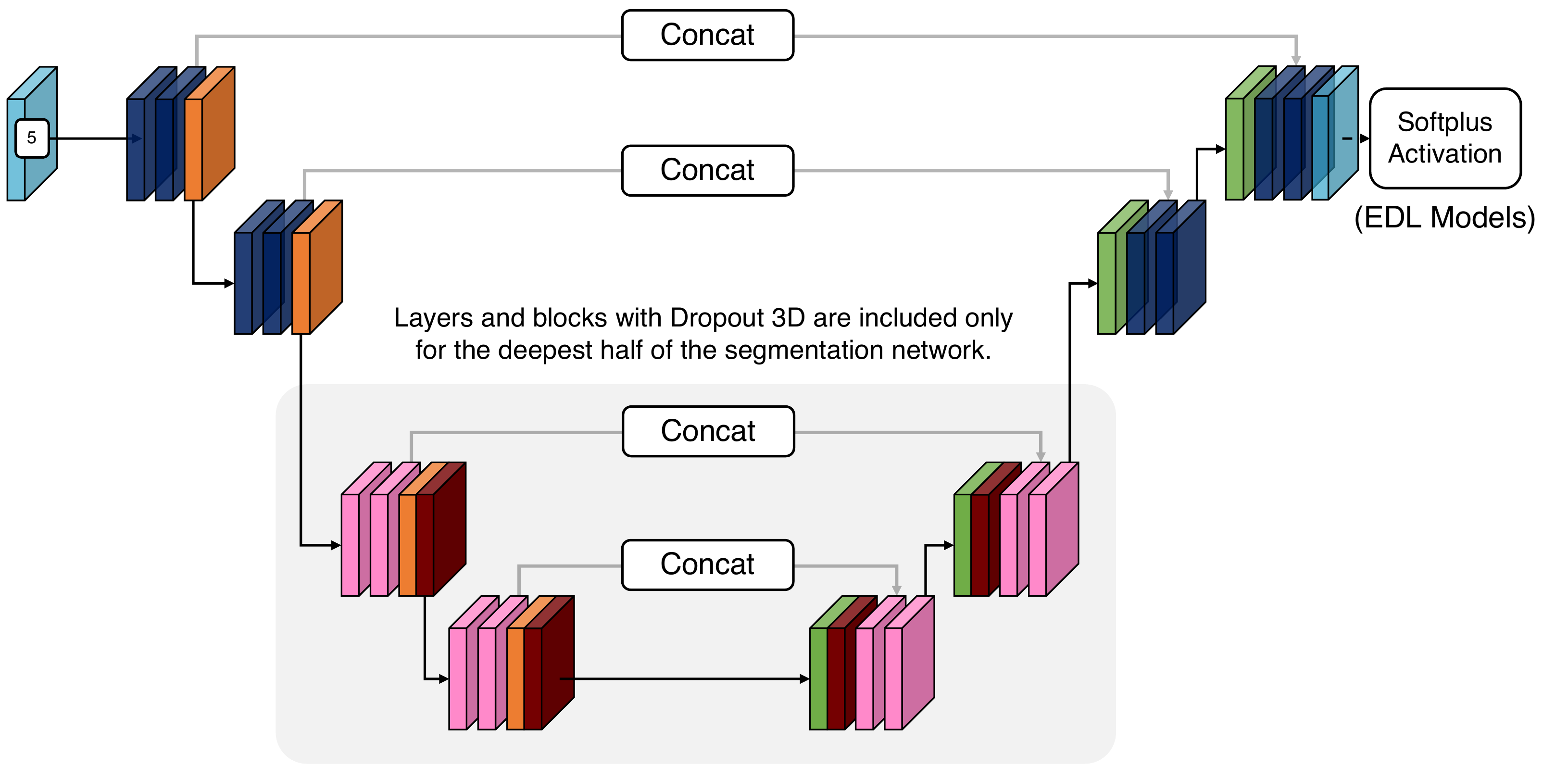}
    \caption{Sparse-CNN architecture for semantic segmentation networks.}
    \label{fig:arch4}
\end{figure*}

\begin{figure*}[htp]
    \centering
    \includegraphics[width=\textwidth]{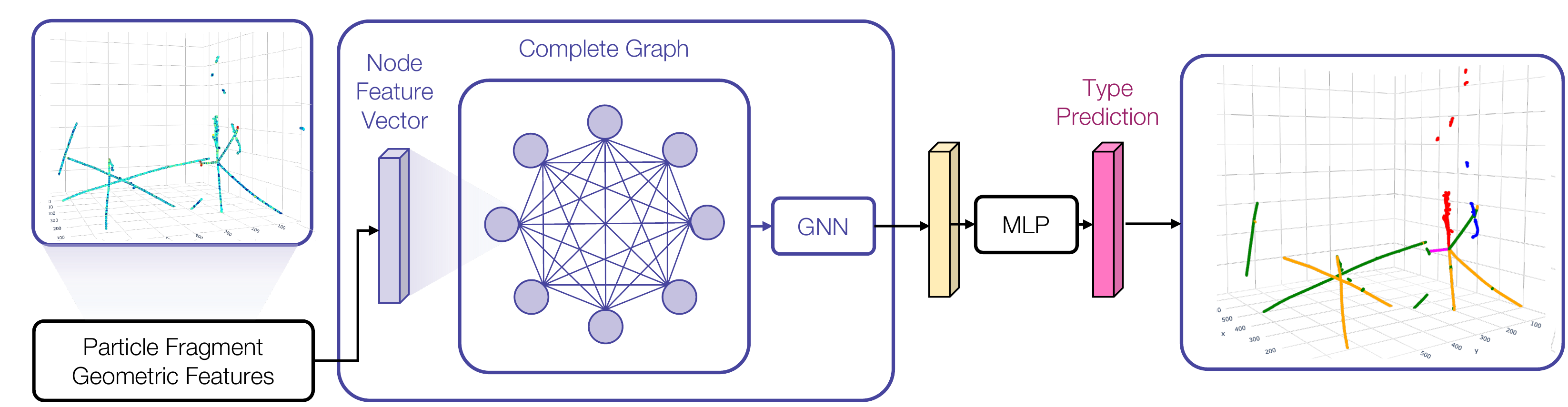}
    \caption{Architecture outline of multi-particle classification network. The geometric node encoder extracts hand-engineered features relevant to particle classification. The colors in the right-most image represent the predicted particle type (blue-photon, red-electron, orange-muon, green-pion, magenta-proton).}
    \label{fig:arch5}
\end{figure*}

\textbf{Network Architecture}: The LArTPC images consist of 3D point clouds, in which after voxelization less than approximately 1\% of the pixels are non-zero. In such a scenario, using standard 3D convolutions over the full grid is not only computationally expensive but also does not process useful information during operation. As such, in 3D applications it is much more common to build convolutional neural network architectures using \textit{sparse convolutions} \cite{mink} that operate solely on active pixels of the full image. Unless stated otherwise, we use MinkowskiEngine's \cite{mink} sparse 3D convolutional layers to build our 3D image processing architectures. 

Figure \ref{fig:arch1} defines the color coding used to represent different architecture components used to describe different models we use to compare UQ methods. Figure \ref{fig:arch2} illustrates definitions for composite blocks that consists of two or more elementary layers in \ref{fig:arch1}. Figures \ref{fig:arch3}, \ref{fig:arch4}, and \ref{fig:arch5} show schematic diagrams of architectures used to evaluate three tasks: single particle classification, particle shape semantic segmentation, and multi particle classification, respectively. 

\textbf{Single Particle Classification}: In the first task, we implement and assess the different UQ models on the simpler task of single particle classification. The single particle dataset consists of 1024 $\times$ 1024 $\times$ 1024 3D images each containing only one particle, where all voxels in the given image belong to the same particle ID. The 3D images have one feature dimension corresponding to the amount of energy deposited in a one-voxel equivalent region of the detector. We use a ResNet \cite{resnet} type encoder with dropout \cite{dropout} regularization, where convolution operations are substituted by sparse convolutions implemented in the \textit{MinkowskiEngine} library \cite{mink}. For standard deterministic models, ensembles, and MCDropout, the final prediction probabilities are given by softmax activations, whereas for evidential models the concentration parameters $\alpha$ are computed from Softplus \cite{softplus} activations. The single particle dataset contains five particle classes: photon showers ($\gamma$), electron showers ($e$), muons ($\mu$), pions ($\pi$), and protons ($p$). A diagram of the neural network architecture is shown in Figure \ref{fig:arch1}.  

\textbf{Semantic Segmentation} As segmentation is a classification task on individual pixels, the details of the implementation are mostly identical to those of single particle classification. We employ a \textit{Sparse-UResNet} \cite{laura} with dropout layers in the deeper half of the network as the base architecture for semantic segmentation networks and use the 768px resolution PILArNet \cite{pilarnet} MultiPartRain (MPR) and MultiPartVertex (MPV) datasets for multiple particle datasets. The five semantic labels provided by PiLArNet consists of the following:
\begin{itemize}
    \item Shower Fragments: connected components of electromagnetic showers that are above a set voxel count and energy deposition threshold. 
    
    \item Tracks: particle trajectories that resemble straight lines, mostly originating from muon, pion, and protons. 
    
    \item Michel Electrons: an electron produced by muon decay at rest.
    
    \item Delta Rays: electrons produced from muon tracks via hard scattering
    
    \item Low Energy Depositions: clouds of low energy depositions of electromagnetic showers which are not labeled as shower fragments. 
\end{itemize}

\textbf{Multi Particle Reconstruction} The MPV/MPR dataset also contains particle type labels for each particle instance in a given image. For multi particle classification, we take each cluster of voxels that belong to the same particle and reduce the resulting groups of point clouds into 1-dimensional feature vectors. The node embeddings of each particle consists of geometric features such as the trajectory start point, initial direction vector, and mean energy deposition. These feature vectors are then given as input node features to a graph neural network, which performs three message passing operations to incorporate inter-particle relational information. 

\section{Results}

In the ensuing discussions, we use the following abbreviations: Det (Deterministic), NE (Naive Ensembles), BE (Bootstrapped Ensembles), MCD (MCDropout), EDL-MLL (Evidential Deep Learning with Marginal Log-Likelihood), EDL-BR (EDL with Bayes Risk), and EDL-B (EDL with Brier Score) to label each model in Figures \ref{fig:singlep} to \ref{fig:cali_seg}. 

\subsection{Training Details}

\begin{figure*}[htp!]
    \centering
    \includegraphics[width=0.99\textwidth]{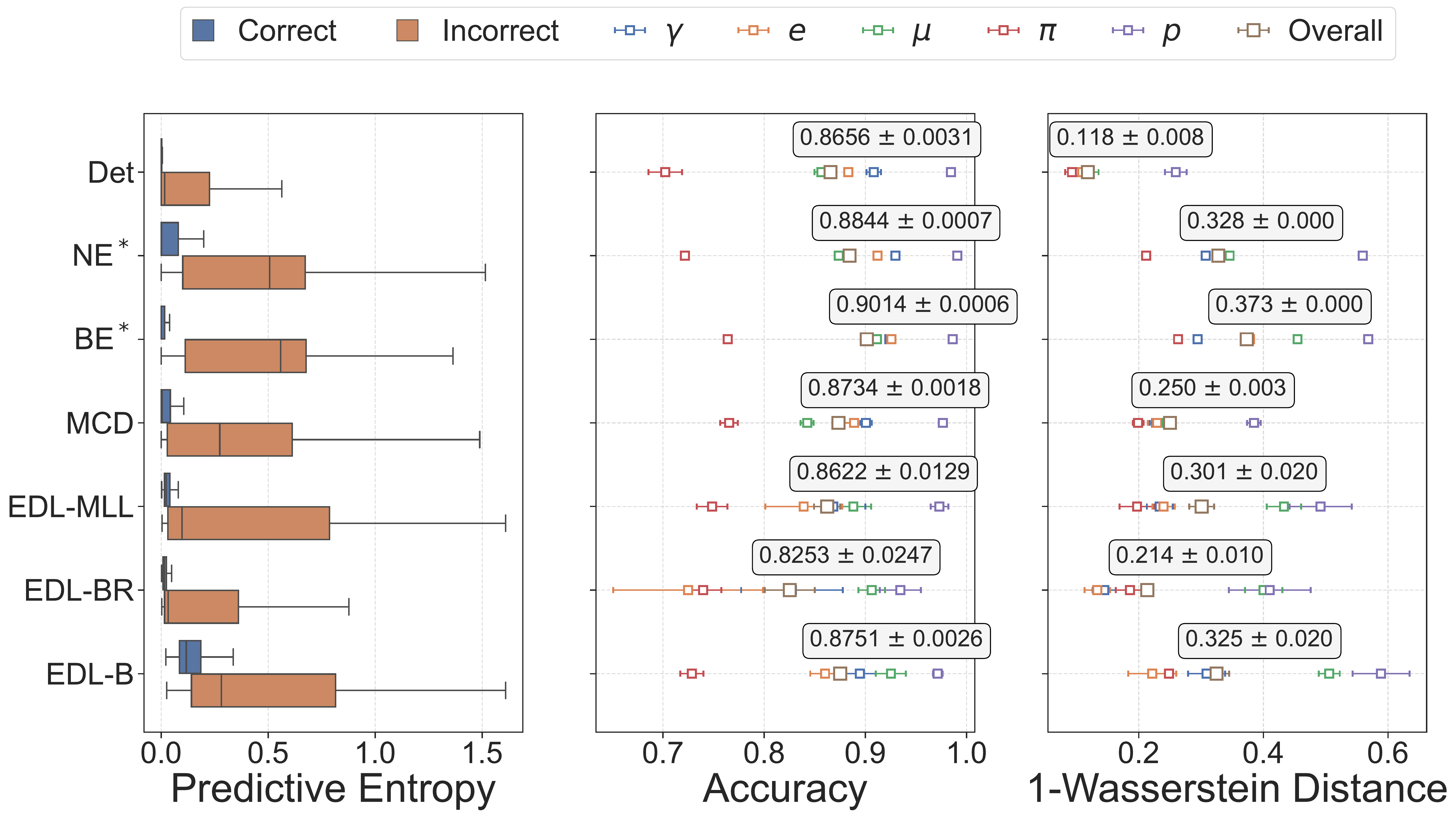}
    \caption{Predictive entropy distribution, accuracy, and 1-Wasserstein distance for single particle classification.}
    \label{fig:singlep}
\end{figure*}

\begin{figure*}[htp!]
    \centering
    \includegraphics[width=0.99\textwidth]{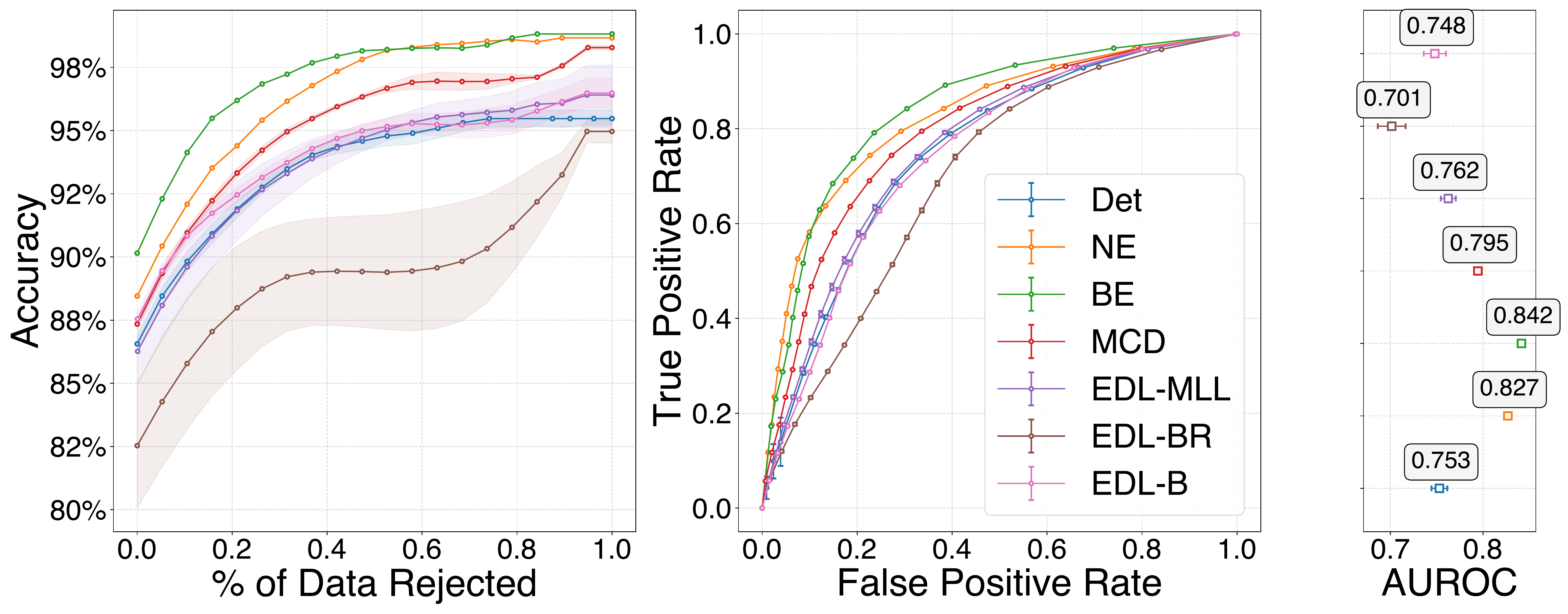}
    \caption{Single particle ROC and percentage rejection curves.}
    \label{fig:roc_singlep}
\end{figure*}

\begin{figure*}[htp!]
    \centering
    \includegraphics[width=0.98\textwidth]{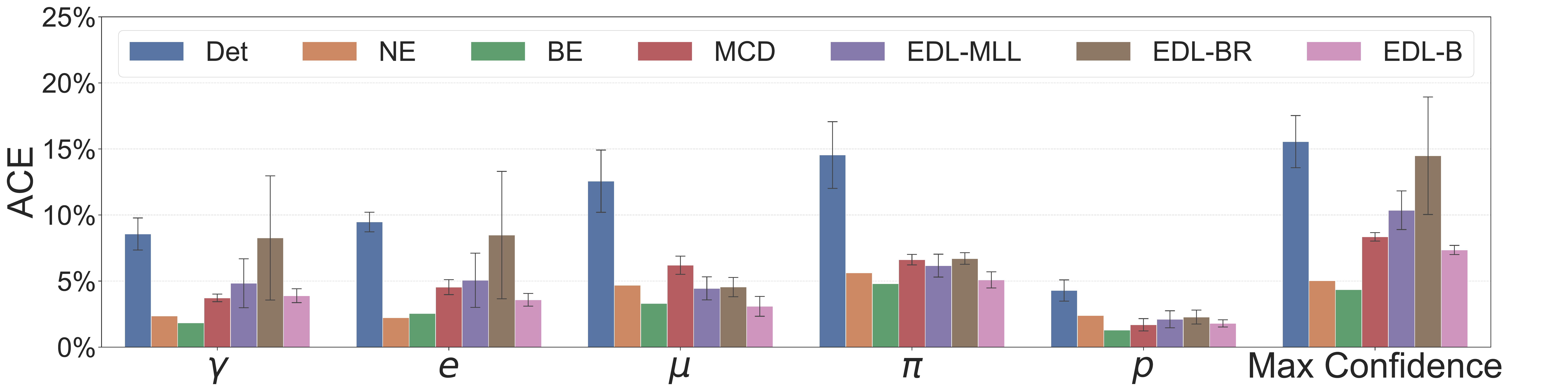}
    \caption{Single particle classification adaptive calibration errors (ACEs) for each model and class.}
    \label{fig:ace_singlep}
\end{figure*}

\begin{figure*}[htp!]
    \centering
    \includegraphics[width=0.99\textwidth]{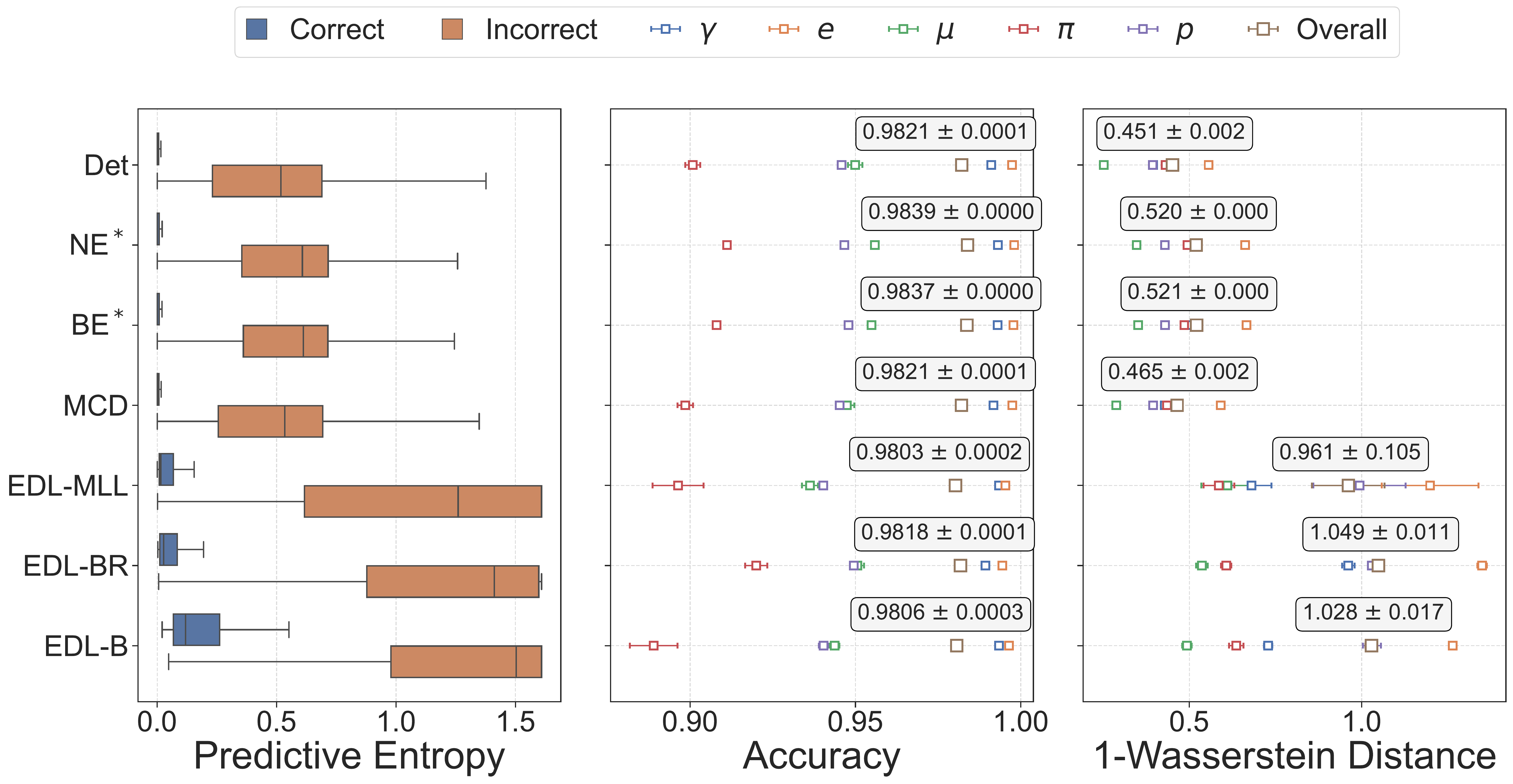}
    \caption{Predictive entropy distribution, accuracy, and 1-Wasserstein distance for semantic segmentation.}
    \label{fig:seg_box}
\end{figure*}

\begin{figure*}[htp!]
    \centering
    \includegraphics[width=\textwidth]{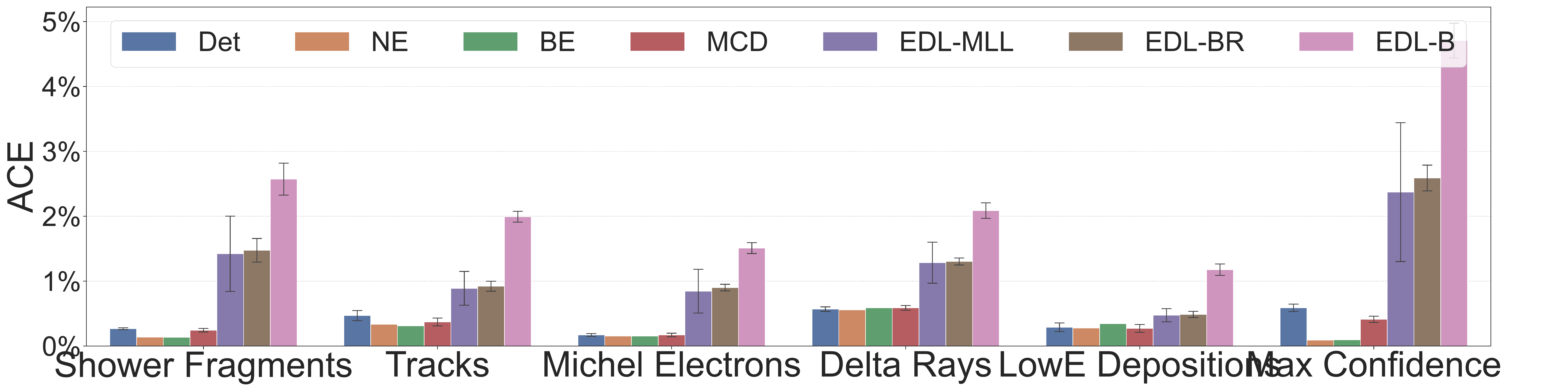}
    \caption{Semantic segmentation adaptive calibration errors (ACEs) for each model and class.}
    \label{fig:ace_seg}
\end{figure*}

\begin{figure*}[htp!]
    \centering
    \includegraphics[width=0.99\textwidth]{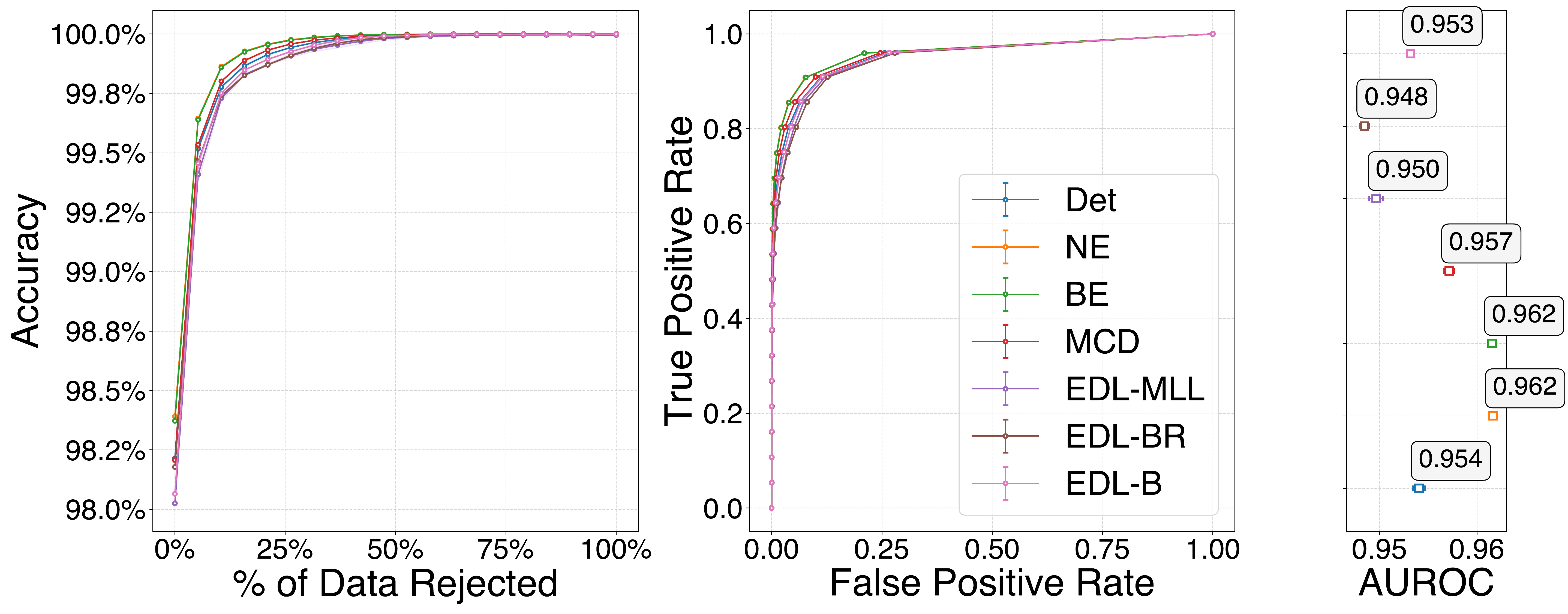}
    \caption{Semantic segmentation ROC and percentage rejection curves.}
    \label{fig:roc_seg}
\end{figure*}

\begin{figure*}[htp!]
    \centering
    \includegraphics[width=0.99\textwidth]{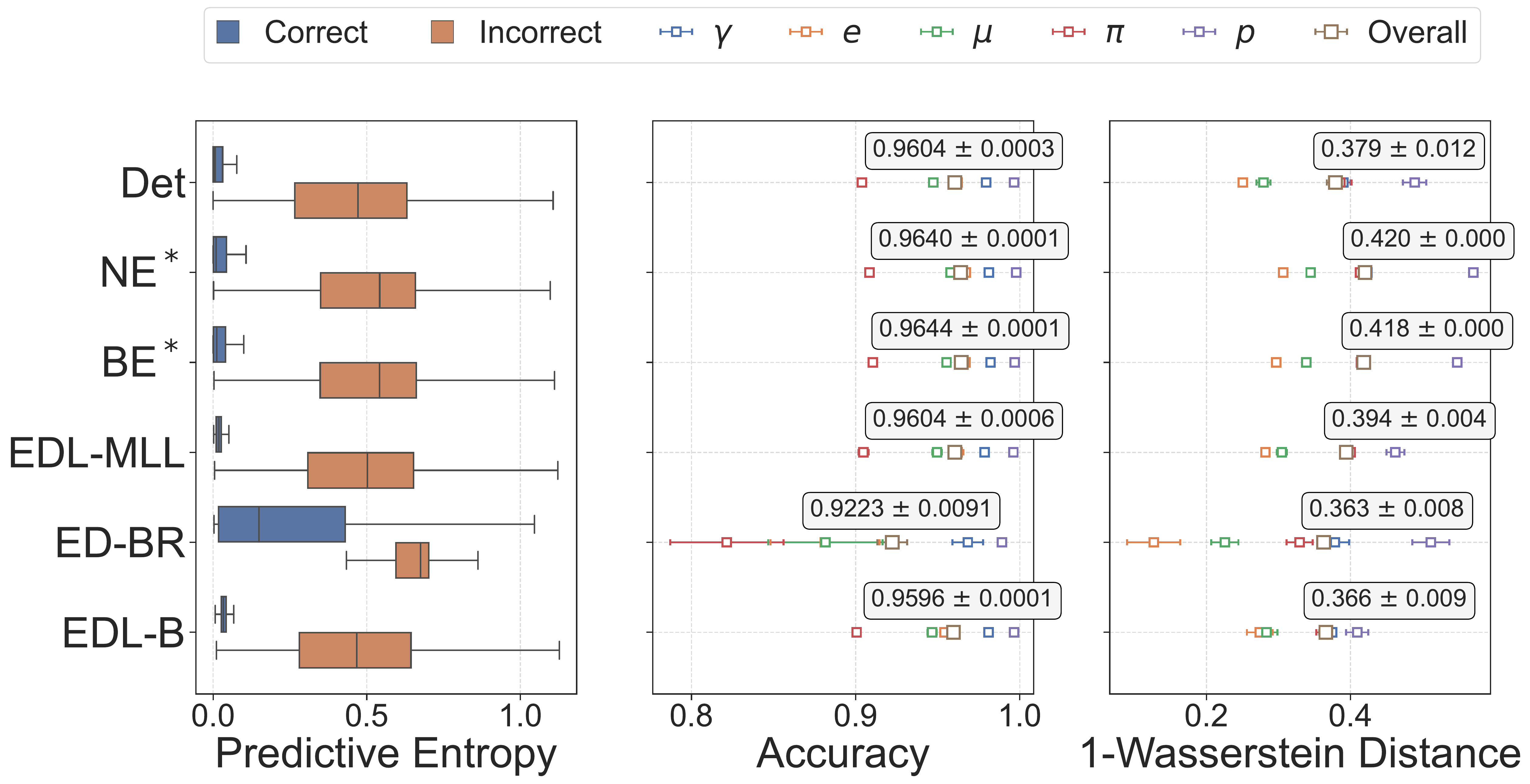}
    \caption{Predictive entropy distribution, accuracy, and 1-Wasserstein distance for multi particle classification.}
    \label{fig:gnn}
\end{figure*}

\begin{figure*}[htp!]
    \centering
    \includegraphics[width=\textwidth]{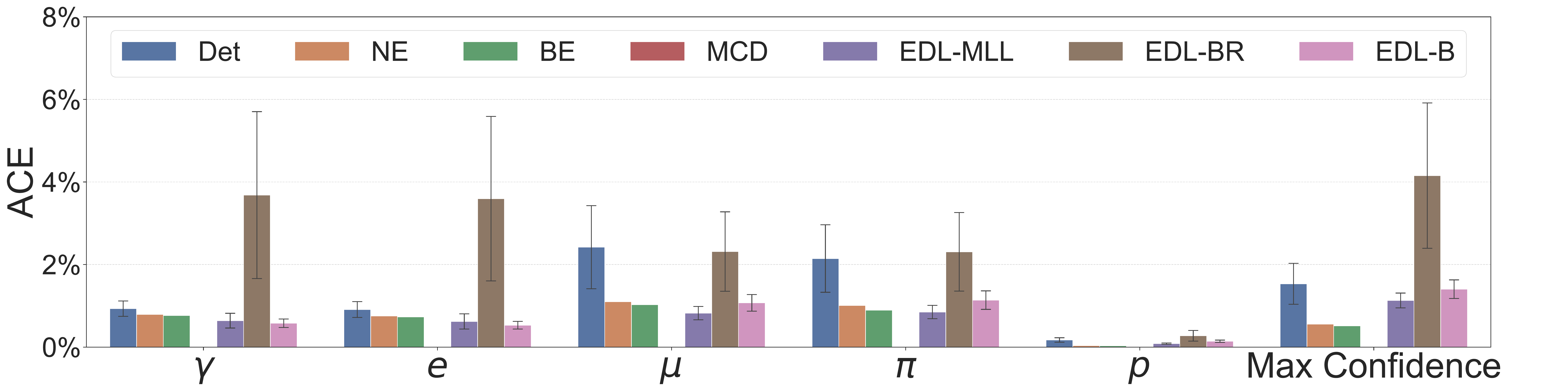}
    \caption{Multi particle classification adaptive calibration errors (ACEs) for each model and class.}
    \label{fig:ace_gnn}
\end{figure*}

\begin{figure*}[htp!]
    \centering
    \includegraphics[width=0.99\textwidth]{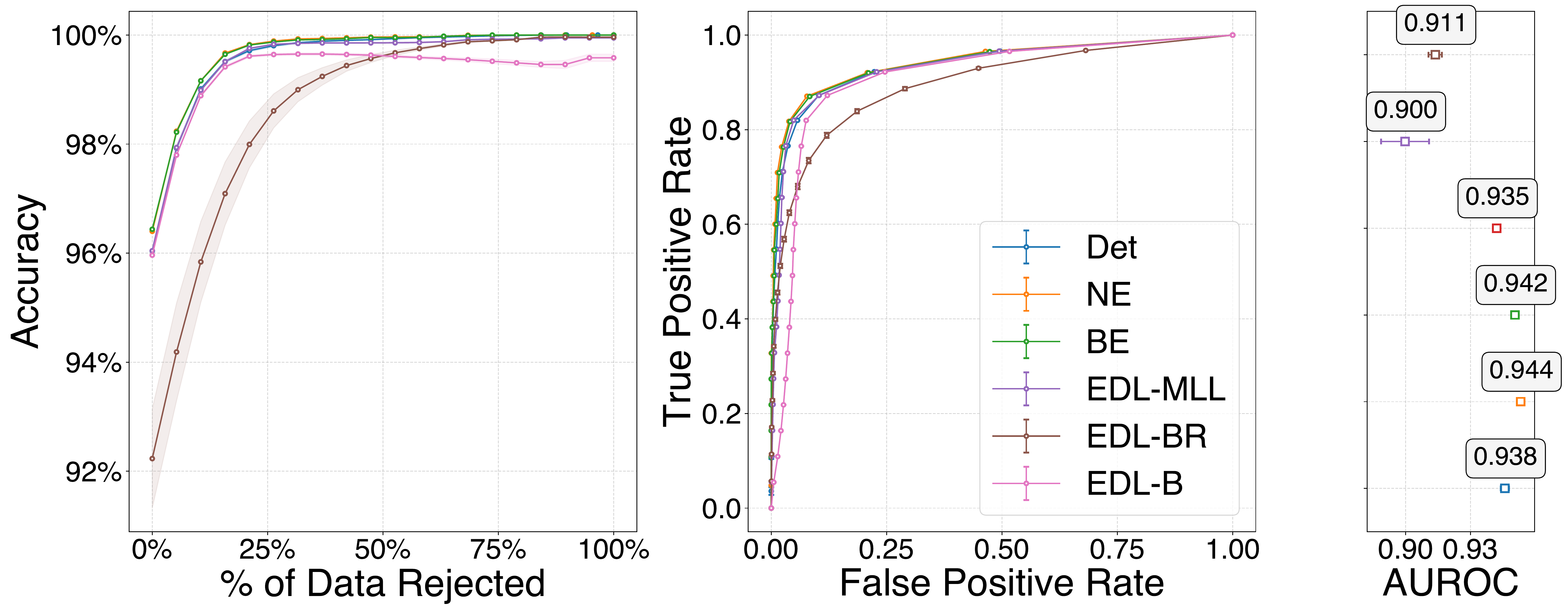}
    \caption{Multi particle ROC and percentage rejection curves.}
    \label{fig:roc_gnn}
\end{figure*}

\begin{figure*}[htp!]
    \centering
    \includegraphics[width=0.99\textwidth]{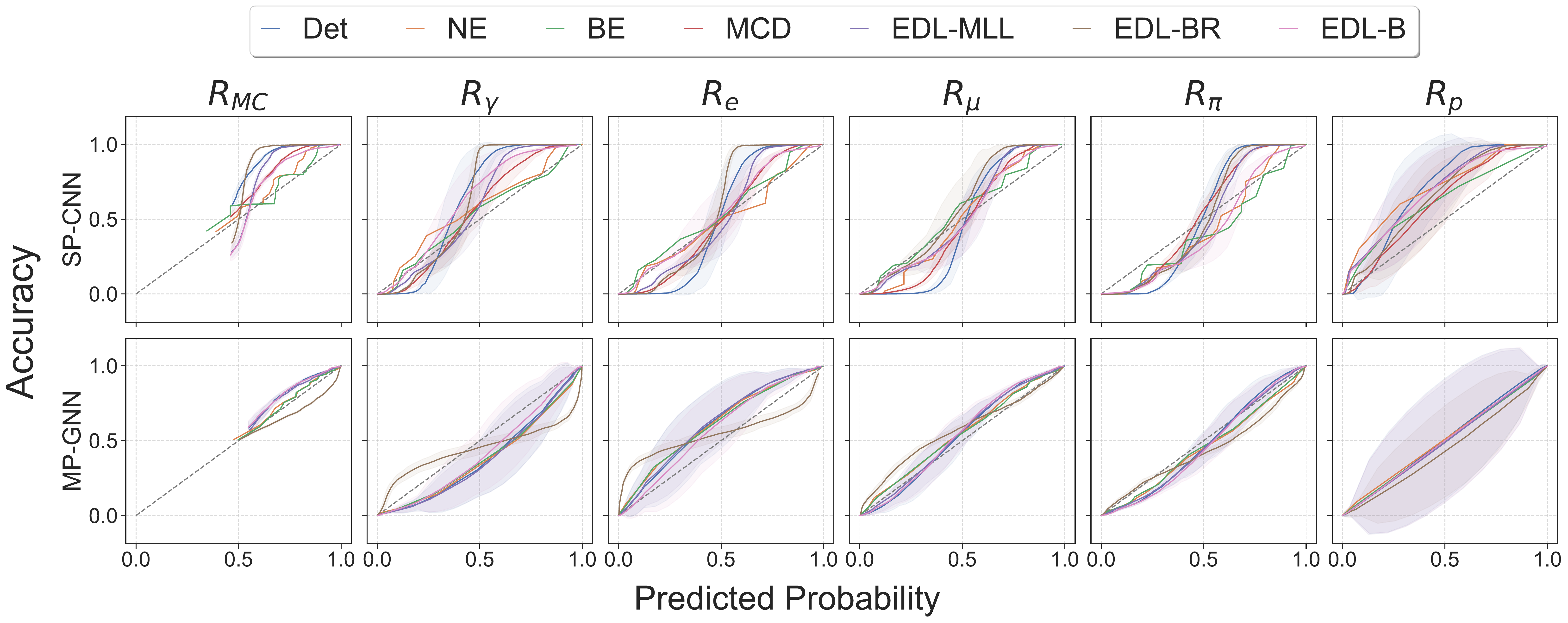}
    \caption{Reliability plots for single and multi-particle classification.}
    \label{fig:cali_particle}
\end{figure*}

\begin{figure*}[htp!]
    \centering
    \includegraphics[width=0.99\textwidth]{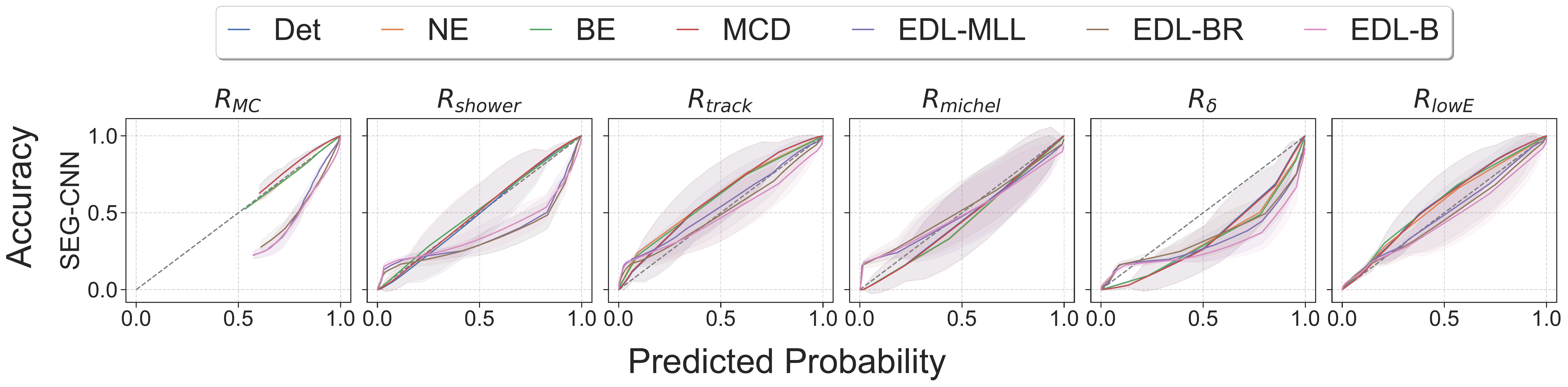}
    \caption{Reliability plots for semantic segmentation.}
    \label{fig:cali_seg}
\end{figure*}

The training set consists of 80k images, and the original test set was separated into a 2k validation set used for model selection and a 18k test set used for selected model evaluation with high statistics. All models were trained until the validation accuracy plateaued, and the network weights that achieved the highest validation accuracy were selected for further evaluation on a separate test set. To fully account for possible variations in model accuracy and uncertainty quantification quality due to randomized factors such as parameter initialization, the model selection procedure was repeated for five different random seeds for each model, except for ensemble methods. This results in five independently trained models that share the same architecture but differing in parameter values. For ensembling methods (NE and BE), the results in Figures \ref{fig:singlep} to \ref{fig:cali_seg} represent predictions from a single ensemble model with five independently trained ensemble members. As such, we mark the results for NE and BE with an asterisk. We used the Adam optimizer \cite{adam} with decoupled weight decay \cite{adamw}.

\textbf{Single Particle Classification}: Figure \ref{fig:singlep} shows the predictive entropy distribution, accuracy, and the $W_1$ distance for the single particle classification models. We observe that the distributional separation as measured in $W_1$ is largest for the ensemble methods, while the Evidential model trained on the Brier score is also competitive. In general, ensemble methods achieve highest accuracy with better distributional separation compared to Monte Carlo Dropout and Evidential models. The AUROC values in figure \ref{fig:roc_singlep} also reflect the superior discriminative capacity of ensembling. 

The calibration curves for single particle classification are shown in the top row of figure \ref{fig:cali_particle}, and figure \ref{fig:ace_singlep} illustrates the adaptive calibration error (ACE) values across different subsets of the test set partitioned by true particle ID labels. While all UQ models with the possible exception of EDL-BR-B achieve better calibration compared to standard deterministic neural networks, ensembling methods have the least max-confidence and marginal ACE values.

\textbf{Semantic Segmentation}: For segmentation, the best distributional separation is achieved by Evidential models, which is exhibited in Figure \ref{fig:seg_box}. The ensemble methods have the highest accuracy and AUROC scores as is shown in figure \ref{fig:ace_seg}. It is interesting to note that while distributional separation measured in $W_1$ is greatest for Evidential models, the calibration fidelity falls short even with respect to standard deterministic models. As with single particle classification, the best calibration fidelity is realized by ensemble methods. 

\textbf{Multi Particle Reconstruction}: Since the contextual information which is useful in determining a given particle's ID can only be used in a multi-particle setting, we expect a gain in accuracy from the single particle datasets. This approach leads to an overall approximate $5\%$ increase in classification accuracy in all models. Again, ensemble methods provide the highest $W_1$ distance, overall accuracy, and AUROC values (figures \ref{fig:gnn}, \ref{fig:roc_gnn}) and the best calibration fidelity (figure \ref{fig:ace_gnn}).

The full reliability plots used to calculate ACE values are provided in figures \ref{fig:cali_particle} and \ref{fig:cali_seg}. A tabular summary of results is available in Table \ref{tab:table3}, where each row corresponds to a different UQ approach and each column corresponds to a different performance metric. We utilize bold font to highlight the best performing approach for a specific performance metric (that is, best performance for the entire  column). Across different tasks and evaluation metrics, we see that ensembling methods provide the best performance. Of these, we find that Bootstrapped Ensembles exhibit superior performance in most cases. Additionally, we highlight the fact  that for some approaches such as EDL-BR, the performance of the uncertainty enabled model is inferior to the baseline deterministic model. For instance, EDL-BR is not as accurate as the deterministic model for all three test cases of Semantic Segmentation, Single-Particle and Multi-Particle Classification. Similarly, both EDL-Brier and EDL-BR exhibit inferior AUROC scores than the deterministic approach across the test cases. 


\begin{landscape}
\mbox{}\vfill
\begin{center}
\begin{table}[t]
\caption{Uncertainty Quantification Evaluation Metrics.}
\begin{tabular}{rccccccccccccccccccccccc}  
\toprule
& \multicolumn{5}{c}{Single Particle Classification} & \multicolumn{5}{c}{Semantic Segmentation} &  \multicolumn{5}{c}{Multi-Particle Classification}\\
\cmidrule(r){2-6} \cmidrule(r){7-11} \cmidrule(r){12-16}
& {Accuracy} & {AUROC} & {$W_1$} & ECE & & {Accuracy} & {AUROC} & {$W_1$} & ECE & & {Accuracy} & {AUROC} & {$W_1$} & ECE & \\
\midrule
Deterministic & {0.8656} & {0.753} & {0.118} & {15.55\%} & & {0.9821} & {0.954} & {0.451} & {0.59\%} & & {0.9604} & {0.938} & {0.379} & {1.53\%}\\
Naive Ensembles & {0.8844} & {0.827} & {0.328} & {5.02\%} & & {\textbf{0.9839}} & {\textbf{0.962}} & {0.520} & {\textbf{0.09\%}} & & {0.9640} & {\textbf{0.944}} & {\textbf{0.420}} & {\textbf{0.56\%}}\\
Bootstrap Ensembles & {\textbf{0.9014}} & {\textbf{0.842}} & {\textbf{0.373}} & {\textbf{4.35\%}} & & {0.9837} & {\textbf{0.962}} & {\textbf{0.521}} & {0.10\%} & & {\textbf{0.9644}} & {0.942} & {0.418} & {0.51\%}\\
MC Dropout & {0.8734} & {0.795} & {0.250} & {8.35\%} & & {0.9821} & {0.957} & {0.465} & {0.41\%} & & {--} & {--} & {--} & {--}\\
EDL-MLL & {0.8622} & {0.762} & {0.301} & {10.36\%} & & {0.9803} & {0.950} & {0.961} & {2.37\%} & & {0.9604} & {0.935} & {0.394} & {1.13\%}\\
EDL-BR & {0.8253} & {0.701} & {0.214} & {14.49\%} & & {0.9818} & {0.948} & {1.049} & {2.59\%} & & {0.9223} & {0.900} & {0.363} & {4.15\%}\\
EDL-Brier & {0.8751} & {0.748} & {0.325} & {7.36\%} & & {0.9806} & {0.953} & {1.028} & {4.70\%} & & {0.9596} & {0.911} & {0.366} & {1.40\%}\\
\bottomrule
\end{tabular}
\label{tab:table3}
\end{table}
\mbox{}\vfill
\begin{table}[t]
    \caption{Average time spent per iteration (seconds).}
    \begin{tabular}{rcccccccccccccc}  
    \toprule
    & \multicolumn{2}{c}{Single Particle Classification} & & \multicolumn{2}{c}{Semantic Segmentation} & & \multicolumn{2}{c}{Multi-Particle Classification}\\
    \cmidrule(r){2-4} \cmidrule(r){5-7} \cmidrule(r){8-9}
    & Train (256) & Test (100) & & Train (64) & Test (100) & & Train (128) & Test (100) \\
    \midrule
    Deterministic & {0.873} & {0.098} & & {1.047} & {0.211} & & {5.056} & {3.284}\\
    Naive Ensembles & {0.873 ($\times 5$)} & {0.098 ($\times 5$)} & & {1.047 ($\times 5$)} & {0.211 ($\times 5$)} & & {5.056 ($\times 5$)} & {3.284 ($\times 5$)}\\
    Bootstrap Ensembles & {4.84 ($\times 5$)} & {0.096 ($\times 5$)} & & {1.386 ($\times 5$)} & {0.176 ($\times 5$)} & & {6.587 ($\times 5$)} & {2.993 ($\times 5$)}\\
    MC Dropout & {0.873} & {2.581} & & {1.047} & {2.049} & & {--} & {--}\\
    EDL-MLL & {0.776} & {0.067} & & {0.964} & {0.235} & & {5.650} & {2.899}\\
    EDL-BR & {0.410} & {0.088} & & {1.058} & {0.280} & & {5.280} & {3.610}\\
    EDL-Brier & {0.790} & {0.089} & & {1.114} & {0.276} & & {5.918} & {3.602}\\
    \bottomrule
    \end{tabular}
    \label{tab:table_time}
\end{table}
\mbox{}\vfill
\begin{table}[t]
    \caption{Average memory consumption per iteration (GBs).}
    \begin{tabular}{rcccccccccccccc}  
    \toprule
    & \multicolumn{2}{c}{Single Particle Classification} & & \multicolumn{2}{c}{Semantic Segmentation} & & \multicolumn{2}{c}{Multi-Particle Classification}\\
    \cmidrule(r){2-4} \cmidrule(r){5-7} \cmidrule(r){8-9}
    & Train (256) & Test (100) & & Train (64) & Test (100) & & Train (128) & Test (100) \\
    \midrule
    Deterministic & {4.245} & {0.66} & & {3.26} & {0.90} & & {0.65} & {0.11}\\
    Naive Ensembles & {4.245 ($\times 5$)} & {0.66 ($\times 5$)} & & {3.26 ($\times 5$)} & {0.90 ($\times 5$)} & & {0.65 ($\times 5$)} & {0.11 ($\times 5$)}\\
    Bootstrap Ensembles & {4.454 ($\times 5$)} & {0.66 ($\times 5$)} & & {3.42 ($\times 5$)} & {0.90 ($\times 5$)} & & {0.65 ($\times 5$)} & {0.11 ($\times 5$)}\\
    MC Dropout & {4.245} & {0.66} & & {3.26} & {0.95} & & {-} & {-}\\
    EDL-MLL & {4.250} & {0.66} & & {3.31} & {0.90} & & {0.65} & {0.11}\\
    EDL-BR & {4.260} & {0.66} & & {3.31} & {0.90} & & {0.65} & {0.11}\\
    EDL-Brier & {4.249} & {0.66} & & {3.33} & {0.90} & & {0.65} & {0.11}\\
    \bottomrule
    \end{tabular}
    \label{tab:table_mem}
\end{table}
\end{center}
\end{landscape}

We include a brief summary of the required time complexity and GPU memory requirements of each model in tables \ref{tab:table_time} and \ref{tab:table_mem} respectively. The batch size is denoted in the Train/Test sub-column. The time information corresponds to the CPU time it takes for the model to run one iteration of training or evaluation routine with the denoted batch size. For training, this includes the time required for both model forwarding and gradient backpropagation, while for inference we compute the sum of evaluation mode model forwarding time and other post-processing operations (for example, in MC Dropout we have a sample averaging procedure needed to obtain probability values). The memory value is computed by taking the average of the maximum required GPU memory across 5 different samples. Note that the values for deterministic and naive ensembles are identical, since naive ensembles were constructed from trained deterministic models. 

Table \ref{tab:table_mem} shows that memory consumption for each model is roughly uniform across different models, with the exception of ensembling methods which naturally require multiple independent forward operations ($N = 5$) for each prediction. While it is possible to run the ensembling methods as fast as single network (non-ensembling) models by parallelization, for computationally lightweight systems this may not be feasible. Similarly, the average run times shown in Table \ref{tab:table_time} also have the same order of magnitude across different models except for naive and bootstrapped ensembles that scale up with the number of constituent models in the ensemble.

\begin{figure*}[htp!]
    \centering
    \includegraphics[width=0.99\textwidth]{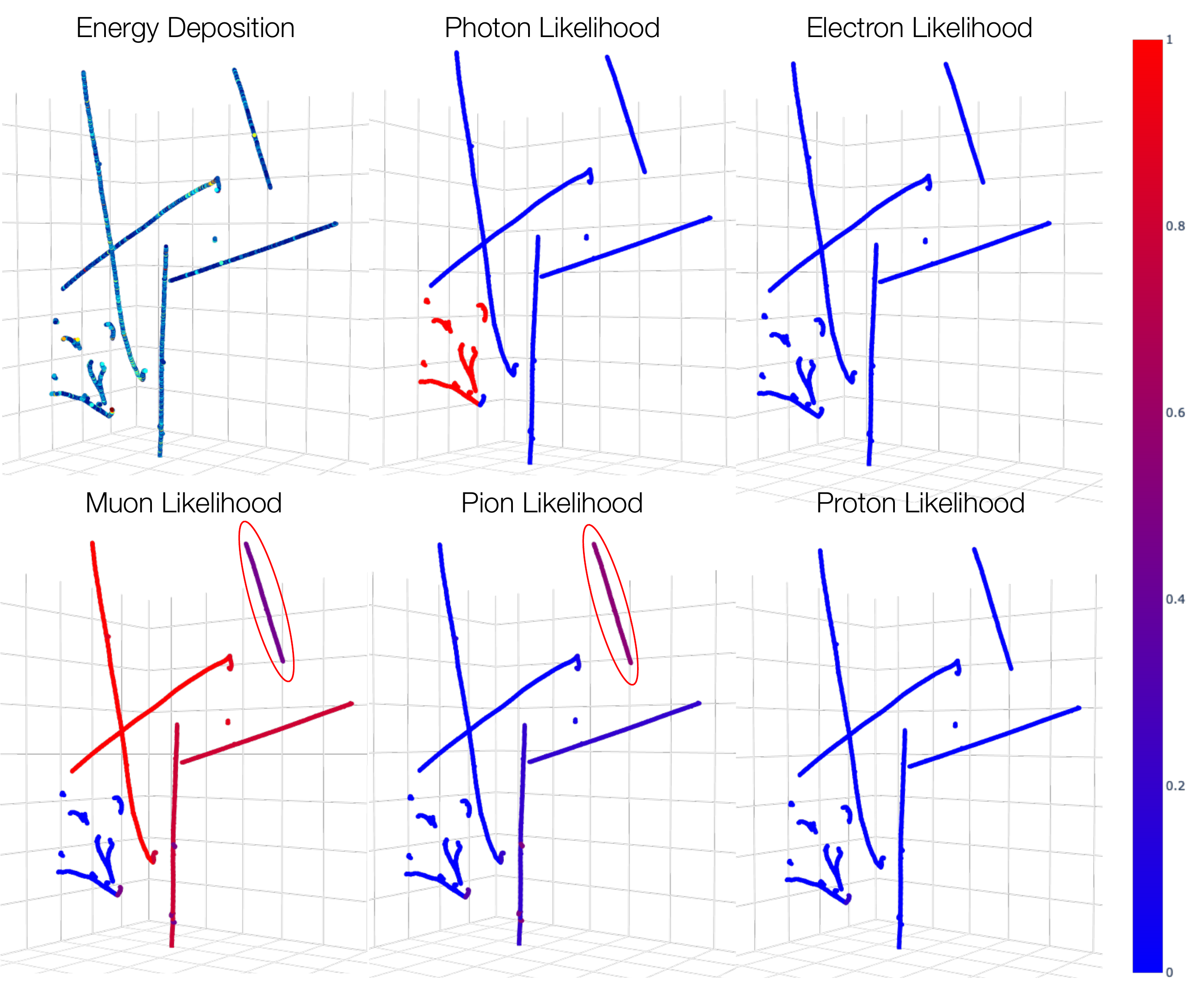}
    \caption{Example high-entropy error from a multi-particle evidential GNN (EDL-BR).}
    \label{fig:ex1}
\end{figure*}
\begin{figure*}[htp!]
    \centering
    \includegraphics[width=0.99\textwidth]{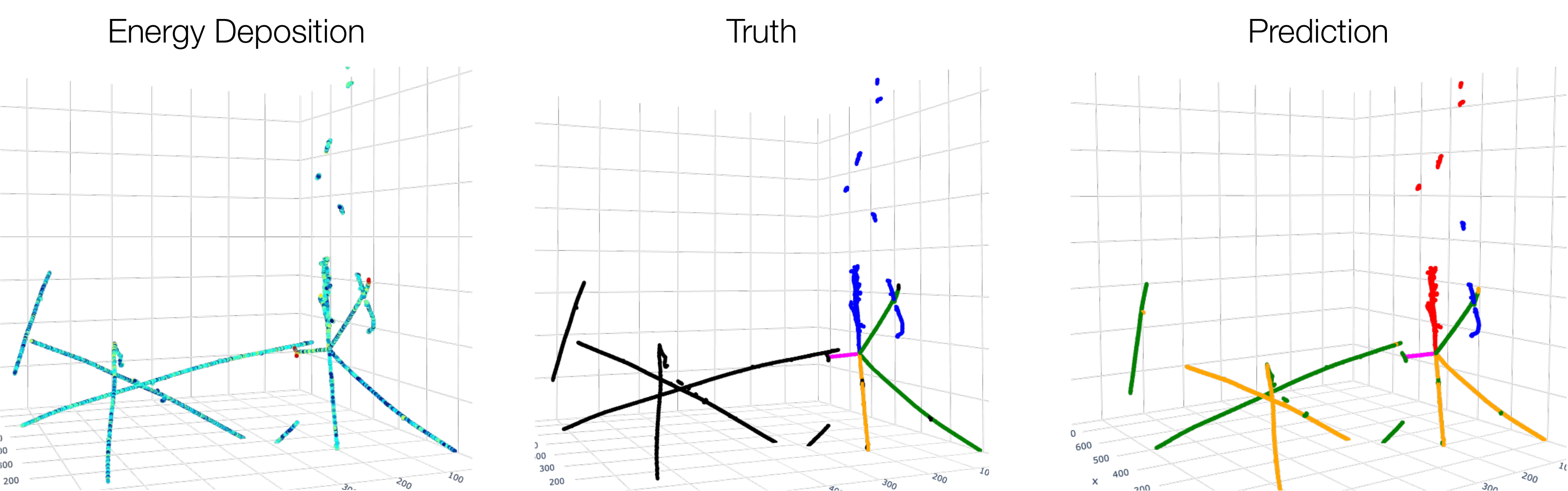}
    \caption{Example low-entropy prediction from a multi-particle evidential GNN (EDL-BR). (blue-photon, red-electron, orange-muon, green-pion, magenta-proton) }
    \label{fig:ex2}
\end{figure*}
\begin{figure*}[htp!]
    \centering
    \includegraphics[width=0.99\textwidth]{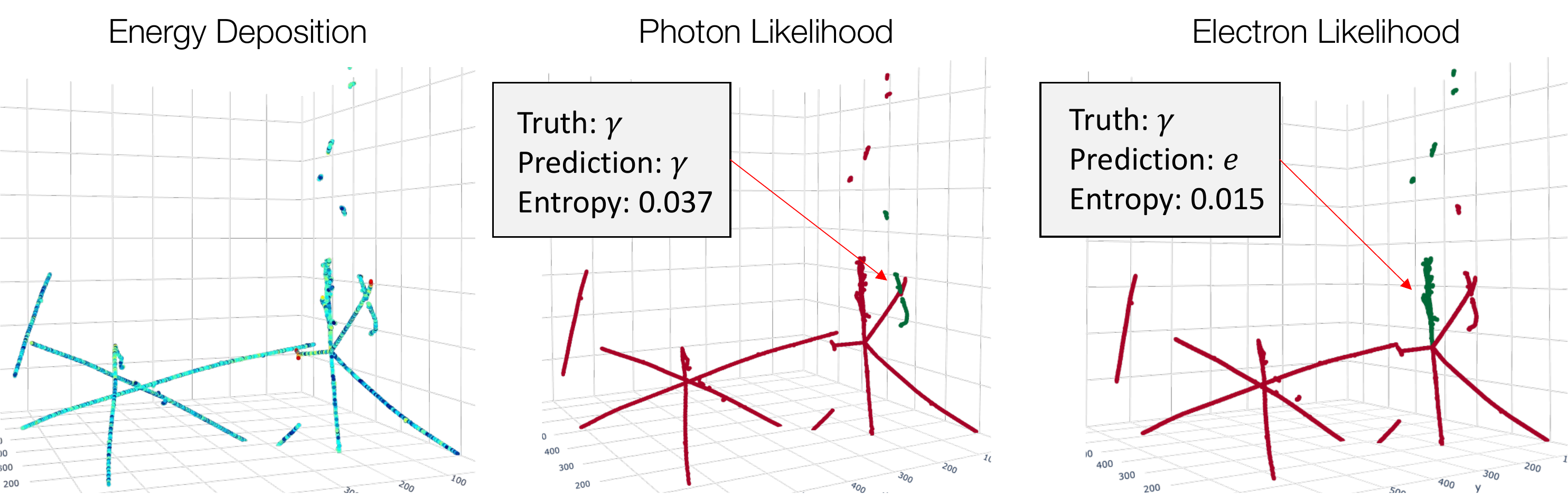}
    \caption{Example low-entropy prediction from a multi-particle evidential GNN (EDL-BR). Green and red indicate high and low probability values, respectively.}
    \label{fig:ex3}
\end{figure*}

Calibration fidelity cannot be examined in a single image basis, as calibration is a collective property that must be measured by appropriate binning of the test set predictions. However, it is possible to assess the discriminative capacity by observing samples with antipodal entropy values. With predictive entropy values in hand, the class predictions may be divided into four categories: 1) confident (low entropy) correct predictions, 2) uncertain (high entropy) correct predictions, 3) confident errors, and 4) uncertain errors. Among the four groups, confident errors are most problematic for robust design of deep learning models. Some representative examples are shown in figures \ref{fig:ex1}, \ref{fig:ex2}, and \ref{fig:ex3}. Figure \ref{fig:ex1} is a high-entropy misclassification example, in which the network cannot confidently decide whether the set of voxels circled in red is a muon or a pion, in contrast with the confident predictions it gives for the pair of photons in and the vertex-attached proton. It is likely that the model cannot determine the circled track as a muon or a pion with certainty, since no inter-particle information such as michel electrons or pion secondary interactions are available for the circled track. In figures \ref{fig:ex2} and \ref{fig:ex3}, the network predicts the vertex-attached shower as an electron with high probability, while for the two $\mu$ and $\pi$ pair it retains some level of uncertainty. Hence, we observe that the assessment of the network on mis-identifying the shower as an electron is partly justified, as it is difficult to distinguish a photon shower attached to an interaction vertex from an electron shower. 

\section{Discussion}

In this investigation, we evaluated three different uncertainty quantification methods for deep neural networks on the task of single particle classification, multi-particle classification, and semantic segmentation using high resolution 3D LArTPC energy deposition images. The various metrics evaluating calibration fidelity and discriminative capacity lead to a notable conclusion: simple ensembling of few independently trained neural networks generally achieves highest accuracy and best calibration of output probability values. Also, we observe that the quality of uncertainty quantification depends greatly on the type of the classifier's task, and it is possible for Bayesian models to perform worse than deterministic networks in calibration. 

Often, the choices made in hyperparameters and neural network architecture significantly affect the classifier's capacity to achieve the desired performance. It is important to note that the UQ methods presented in this paper do not consider the effects of the \textit{structure} and \textit{hyperparameters} of our models. Extant deep learning uncertainty quantification approaches can only account for aleatoric uncertainty and the parameter uncertainty component of epistemic uncertainty. Thus, these methods are unable to account for structural (or model form) uncertainty, that is a component of epistemic uncertainty. While a complete description of epistemic uncertainty is often intractable in practice, it is desirable to assess how much of the variability in a deep classifier's predictions could be attributed to hyperparameter and structural diversity. 

Additionally, most deep learning uncertainty quantification approaches introduce hyperparameters regarding their usage that need to be selected. The performance of the uncertainty estimation is sensitive to these choices. For instance, in the application of MCDropout, the hyperparameters include the locations of the dropout layers in the network and the dropout probabilities of each layer. Prior investigations have observed that the uncertainty quantification capabilities of this approach are highly affected by these choices, where for instance, choosing too many dropout layers or too strong a dropout probability can lead to overly conservative estimates \cite{kendall2015bayesian}. Similarly, for ensembling approaches, number of constituent models in the ensemble needs to be ascertained; in variation Bayes approaches the scaling between the negative-log-likelihood and the KL-divergence losses needs to be estimated for best performance \cite{mishra2021uncertainty}, etc. In this investigation, we carried out such hyperparameter selection using manual exploration over a validation dataset.

While out-of-distribution and mis-classification resilience of uncertainty quantifying neural networks may be used for rejecting unreliable predictions, obtaining calibrated probability estimates would provide further credibility in using deep learning techniques for physical sciences. Post-hoc calibration methods such as temperature scaling \cite{guo_calibration} train a calibration model (for temperature scaling, a single parameter calibration model) after training to obtain calibrated probabilities from a deterministic neural network. As post-hoc methods do not require the classifier to be re-modeled and trained from their initial state, such methods may be better suited for ensuring proper calibration of classifiers with lower computational budget. Future work will include evaluation of uncertainty quantifying neural networks and post-hoc calibration methods for a full neutrino physics signal/background classifier, which is built on top of the separate tasks of particle classification and segmentation.

\section*{Acknowledgment}
This work was supported in part by funding from Zoox, Inc. This work is supported by the U.S. Department of Energy, Office of Science, Office of High Energy Physics, and Early Career Research Program under Contract DE-AC02-76SF00515.

\bibliographystyle{ieeetr}
\bibliography{references.bib}

\end{document}